\documentclass[aps,pra,amsmath,showpacs,amssymb]{revtex4}

\usepackage{graphicx}
\usepackage{color}

\newcommand{\be}{\begin{equation}}
\newcommand{\ee}{\end{equation}}

\newcommand{\bex}{\begin{eqnarray}}
\newcommand{\eex}{\end{eqnarray}}

\begin{document}

\title{No-signaling, intractability and entanglement}
\author{R. Srikanth}
\email{srik@poornaprajna.org}
\affiliation{Poornaprajna Institute of Scientific Research,
Sadashivnagar, Bangalore- 560080, India.}
\affiliation{Raman Research Institute, Sadashivnagar,
Bangalore- 560080, India.}

\begin{abstract}
We consider  the problem of  deriving the no-signaling  condition from
the assumption that, as  seen from a complexity theoretic perspective,
the universe is not an exponential place. A fact that disallows such a
derivation is  the existence  of {\em polynomial  superluminal} gates,
hypothetical primitive  operations that enable  superluminal signaling
but not the efficient  solution of intractable problems.  It therefore
follows, if  this assumption is  a basic principle of  physics, either
that it  must be supplemented with additional  assumptions to prohibit
such  gates, or,  improbably,  that no-signaling  is  not a  universal
condition.  Yet, a gate of  this kind is possibly implicit, though not
recognized  as  such,  in  a  decade-old  quantum  optical  experiment
involving  position-momentum entangled  photons.  Here  we  describe a
feasible  modified  version  experiment  that  appears  to  explicitly
demonstrate the  action of this gate. Some  obvious counter-claims are
shown to  be invalid.  We  believe that the unexpected  possibility of
polynomial  superluminal operations  arises  because some  practically
measured quantum  optical quantities  are not describable  as standard
quantum mechanical observables.
\end{abstract}
\maketitle \date{}

\section{Introduction}
In a multipartite quantum  system, any completely positive map applied
locally to  one part does not  affect the reduced  density operator of
the  remaining  part.   This  fundamental  no-go  result,  called  the
``no-signalling theorem'' implies that quantum entanglement \cite{epr}
does not enable nonlocal (``superluminal'') signaling \cite{nos} under
standard operations,  and is thus consistent  with relativity, inspite
of   the    counterintuitive,   stronger-than-classical   correlations
\cite{bell}   that   entanglement   enables.   For   simple   systems,
no-signaling  follows from  non-contextuality, the  property  that the
probability  assigned to projector  $\Pi_x$, given  by the  Born rule,
Tr$(\rho\Pi_x)$, where $\rho$ is the density operator, does not depend
on  how the  orthonormal  basis set  is completed  \cite{gle57,ash95}.
No-signaling  has also  been treated  as a  basic postulate  to derive
quantum theory \cite{gis01halv}.

It is of interest to  consider the question of whether/how computation
theory,  in particular intractability  and uncomputability,  matter to
the  foundations of (quantum)  physics. Such  a study,  if successful,
could  potentially  allow  us  to   reduce  the  laws  of  physics  to
mathematical  theorems about  algorithms and  thus shed  new  light on
certain  conceptual  issues.   For   example,  it  could  explain  why
stronger-than-quantum   correlations    that   are   compatible   with
no-signaling  \cite{bra06} are disallowed  in quantum  mechanics.  One
strand of thought  leading to the present work,  earlier considered by
us  in Ref.  \cite{sri06},  was the  proposition that  the measurement
problem is  a consequence of basic algorithmic  limitations imposed on
the computational power that can be supported by physical laws. In the
present work,  we would like to  see whether no-signaling  can also be
explained  in  a  similar  way, starting  from  computation  theoretic
assumptions.

The Turing machine (TM) represents an abstraction of the principles of
mechanical computation. The machine consists of a head and a tape. The
head  is capable  of being  in one  of a  finite number  of ``internal
states" and  can read and overwrite  a symbol from a  finite set, then
shifting one block left or right  along the tape. It contains a finite
internal program that directs  its operations.  The central problem in
computer science  is the conjecture that  two computational complexity
classes, {\bf  P} and  {\bf NP}, are  distinct in the  standard Turing
model  of computation.  ${\bf P}$  is the  class of  decision problems
solvable in  polynomial time by  a (deterministic) TM.  ${\bf  NP}$ is
the class  of decision problems  whose solution(s) can be  verified in
polynomial time  by a deterministic TM.   $\#{\bf P}$ is  the class of
counting problems associated with the decision problems in ${\bf NP}$.
The word  ``complete" following a  class denotes a problem  $X$ within
the class, which is maximally hard in the sense that any other problem
in the  class can be  solved in poly-time  using an oracle  giving the
solutions of  $X$ in a  single clock cycle.  For  example, determining
whether  a Boolean  forumla  is satisfied  is  {\bf NP}-complete,  and
counting   the   number    of   Boolean   satisfactions   is   $\#{\bf
P}$-complete. The word ``hard" following a class denotes a problem not
necessarily  in the  class, but  to which  all problems  in  the class
reduce in poly-time.

{\bf P} is often taken to be the class of computational problems which
are  ``efficiently solvable"  (i.e., solvable  in polynomial  time) or
``tractable", although  there are potentially larger  classes that are
considered tractable such as {\bf  RP} \cite{srirp} and {\bf BQP}, the
latter being the class of  decision problems efficiently solvable by a
quantum  computer \cite{srirp}.   {\bf  NP}-complete and  potentially
harder problems, which  are not known to be  efficiently solvable, are
considered intractable in the Turing model.  If ${\bf P} \ne {\bf NP}$
and the universe is a polynomial-- rather than an exponential-- place,
physical  laws cannot  be harnessed  to efficiently  solve intractable
problems, and  {\bf NP}-complete problems  will be intractable  in the
physical world.


That classical physics supports  various implementations of the Turing
machine is  well known. More  generally, we expect  that computational
models supported by a physical  theory will be limited by that theory.
Witten identified  expectation values  in a topological  quantum field
theory with values  of the Jones polynomial that  are $\#{\bf P}$-hard
\cite{wit89}.   There  is  evidence  that  a physical  system  with  a
non-Abelian topological  term in  its Lagrangian may  have observables
that are {\bf NP}-hard, or even \#{\bf P}-hard \cite{mic98}.

Other recent  related works that have studied  the computational power
of  variants  of  standard  physical  theories from  a  complexity  or
computability      perspective      are,      respectively,      Refs.
\cite{cal04,insel,aar05,bram98,sri06}  and  Refs.  \cite{cal04,sri06}.
Ref.  \cite{aar05}  noted that {\bf NP}-complete problems  do not seem
to  be  tractable  using  resources  of  the  physical  universe,  and
suggested that  this might embody a  fundamental principle, christened
the  {\bf  NP}-hardness  assumption  (also  cf.  \cite{sciam}).   Ref.
\cite{srigru}  studies how  insights from  quantum  information theory
could be used to constrain physical laws.  We will informally refer to
the  proposition  that the  universe  is  a  polynomial place  in  the
computational  sense  (to  be  strengthened  below)  as  well  as  the
communication sense by the expression  ``the world is not hard enough"
(WNHE) \cite{sriwnhe}.  In Ref.  \cite{sri06}, we pointed out that the
assumption of  WNHE (and further that  of {\bf P} $\ne$  {\bf NP}) can
potentially   give  a   unified  explanation   of  (a)   the  observed
`insularity-in-theoryspace' of quantum  mechanics (QM), namely that QM
is {\em exactly} unitary,  linear and requires measurements to conform
to the $|\psi|^2$ Born rule \cite{sriw,insel}; (b) the classicality of
the macroscopic world; (c) the lack of quantum physical mechanisms for
non-signaling superquantum correlations \cite{bra06}.

In (a),  the basic idea  is that departure  from one or more  of these
standard  features  of  QM  seems  to invest  quantum  computers  with
super-Turing power to solve hard problems efficiently, thus making the
universe   an  exponential   place,  contrary   to   assumption.   The
possibility (b) arises for the  following reason.  It is proposed that
the WNHE assumption holds not only in the sense that hard problems (in
the  standard  Turing  model)  are  not efficiently  solvable  in  the
physical  world,   but  in  the  stronger  sense   that  any  physical
computation  can be simulated  on a  probabilistic TM  with at  most a
polynomial slowdown  in the number of steps  (the Strong Church-Turing
thesis). Therefore,  the evolution of  any quantum system  computing a
decision problem, could asymptotically be simulated in polynomial time
in the size of  the problem, and thus lies in {\bf  BPP}, the class of
problems  that  can  be  efficiently  solved  by  a  probabilistic  TM
\cite{sribpp}.

Assuming {\bf  BPP} $\ne$  {\bf BQP}, this  suggests that  although at
small  scales,  standard QM  remains  valid  with characteristic  {\bf
  BQP}-like behavior,  at sufficiently large  scales, classical (`{\bf
  BPP}-like') behavior should emerge, and that therefore there must be
a  definite scale-- sometimes  called the  Heisenberg cut--  where the
superposition   principle   breaks    down   \cite{bag00},   so   that
asymptotically, quantum states are  not exponentially long vectors. In
Ref.   \cite{sri06}, we  speculate that  this  scale is  related to  a
discretization of  Hilbert space.   This approach provides  a possible
computation theoretic  resolution to the  quantum measurement problem.
In (c),  the idea  is that  in a polynomial  universe, we  expect that
phenomena in which  a polynomial amount of physical  bits can simulate
exponentially   large   (classical)   correlations,   thereby   making
communication complexity trivial, would be forbidden.

In  the  present work,  we  are  interested  in studying  whether  the
no-signaling theorem follows from the WNHE assumption.  The article is
arranged as follows.   Some results concerning non-standard operations
that  violate  no-signaling  and  help efficiently  solve  intractable
problems,    are   surveyed    in   Sections    \ref{sec:srisum}   and
\ref{sec:sripol},  respectively.  In Section  \ref{sec:sripolynon}, we
introduce   the  concept   of  a   polynomial  superluminal   gate,  a
hypothetical primitive operation that  is prohibited by the assumption
of  no-signaling,   but  allowed  if  instead  we   only  assume  that
intractable problems  should not  be efficiently solvable  by physical
computers.  We examine  the relation between the above  two classes of
non-standard  gates.  We  also describe  an {\em  constant} gate  on a
single qubit or qutrit, possibly the simplest instance of a polynomial
superluminal operation.  A quantum optical realization of the constant
gate, and its application to an experiment involving entangled photons
generated  by  parametric downconversion  in  a  nonlinear crystal  is
presented in Section \ref{sec:ent}. Physicists who could not care less
about  computational complexity  aspects could  skip directly  to this
Section. They may be warned that the intervening sections will involve
mangling QM in  ways that may seem awkward,  and whose consistency is,
unfortunately, not  obvious!  On  the other hand,  computer scientists
unfamiliar with  quantum optics may skip  Section \ref{sec:ent}, which
is  essentially covered in  Section \ref{sec:sriung},  which discusses
quantitative   and   conceptual   issues  surrounding   the   physical
realization of  the constant gate.  Finally, we conclude  with Section
\ref{sec:conclu} by surveying some implications of a possible positive
outcome  of  the proposed  experiment,  and  discussing  how such  an
unexpected physical effect may  fit in with the mathematical structure
of  known  physics.   We   present  a  slightly  abridged  version  of
discussions in this work in Ref. \cite{sriv1}.

\section{Superluminal gates \label{sec:srisum}}

Even minor variants of QM are known to lead to superluminal signaling.
An   example  is   a  variant   incorporating   nonlinear  observables
\cite{ganeshpol91},   unless   the   nonlinearity   is   confined   to
sufficiently small  scales \cite{srisvet,srisvet0}.  In  this Section,
we will review  the case of violation of  no-signling due to departure
from   standard  QM   via   the  introduction   of  (a)   non-complete
Schr\"odinger evolution  or measurement, (b)  nonlinear evolution, (c)
departure from the Born $|\psi|^2$ rule.

In each  case, we  will not  attempt to develop  a non-standard  QM in
detail,  but   instead  content  ourselves   with  considering  simple
representative examples.
 
(a)  {\em  Non-complete  measurements  or  non-complete  Schr\"odinger
  evolution.}    Let  us  consider   a  QM   variant  that   allows  a
non-tracepreserving    (and   hence   non-unitary)    but   invertible
single-qubit operation of the form:
\begin{equation}
\label{eq:g}
G = \left( \begin{array}{ll} 1 & 0 \\ 0 & 1+\epsilon
\end{array}\right),
\end{equation}
where $\epsilon  > 0$ is a  real number.  The  resultant state $\sum_x
\alpha_x  |x\rangle$  must  be   normalized  by  dividing  it  by  the
normalization factor $\sqrt{\sum_x |\alpha_x|^2}$ immediately before a
measurement, making measurements  nonlinear. Given the entangled state
$(1/\sqrt{2})(|01\rangle +  |10\rangle)$ that Alice and  Bob share, to
transmit a  superluminal signal, Alice applies either  $G^m$ (where $m
\geq 1$  is an integer)  or the identity  operation $I$ to  her qubit.
Bob's  particle  is left,  respectively,  in  the state  $\rho_B^{(1)}
\propto                \frac{1}{2}(|0\rangle\langle0|                +
(1+\epsilon)^{2m}|1\rangle\langle1|)$      or      $\rho_B^{(0)}     =
\frac{1}{2}(|0\rangle\langle0|  + |1\rangle\langle1|)$,  which  can in
principle  be distinguished,  the  distance between  the states  being
greater for  larger $m$ (cf.  Section \ref{sec:sripol}),  leading to a
superluminal signal from Alice to Bob.

More generally,  we may  allow non-unitary and  irreversible evolution
but still  conform to no-signaling, provided the  corresponding set of
operator(s) is {\em complete}, i.e., constitutes a partition of unity.
Suppose Alice and  Bob share the state $\rho_{AB}$,  and Alice evolves
her part of $\rho_{AB}$ locally  through the linear operation given by
the set  ${\cal P}$ of  (Kraus) operator elements $\{E_{j}  \equiv e_j
\otimes\mathbb{I}_B,      j=1,2,3,\cdots\}$     \cite{nc00},     where
$\mathbb{I}_B$  is the  identity  operator in  Bob's subspace.   Bob's
reduced   density  operator   $\rho^{\prime}_B$  conditioned   on  her
performing the operation and after normalization is:
\begin{equation}
\label{nosig}
\rho^{\prime}_B          =         {\cal          N}^{-1}         {\rm
  Tr_A}\left[\sum_{j}E_{j}\rho_{AB}E^{\dag}_{j}\right] = {\cal N}^{-1}
    {\rm          Tr}_A\left[\sum_{j}E^{\dag}_{j}E_{j}\rho_{AB}\right],
    ~~~~{\cal                 N}                 =                {\rm
      Tr}_{AB}\left[\sum_{j}E^{\dag}_{j}E_{j}\rho_{AB}\right],
\end{equation} 
where  ${\cal   N}$  is  the  normalization  factor.    We satisfy the
no-signaling condition
$\rho^\prime_B = \rho_B$ only  if $\rho_{AB}$ is unentangled or ${\cal
  P}$ satisfies the completeness relation
\begin{equation}
\sum_j   e_j^\dag  e_j   = \mathbb{I}_A,
\label{eq:srimadhwa}
\end{equation}
which guarantees  that the operation preserves norm  ${\cal N}$.  Here
$\mathbb{I}_A$ is  the identity operator in Alice's  subspace.  In the
above, non-completeness  suffices, and the  nonlinearity introduced by
renormalizing   the   wavefunction   is   not   necessary,   for   the
superluminality. 

If the  system $A$  is subjected to  unitary evolution  or non-unitary
evolution due to noise, or to standard projective measurements or more
general measurements  described by positive  operator valued measures,
the  corresponding   map  satisfies  Eq.    (\ref{eq:srimadhwa}),  and
$\rho_B^{\prime}  = \rho_B$.   For terminological  brevity, we  call a
(non-standard) gate  like $G$, or a non-complete  operation ${\cal P}$
that  enables  superluminal  signaling,  as `superluminal  gate',  and
denote  the set  of  all  superluminal gates  by  `$C^{<}$'.  For  the
purpose of this work, $C^{<}$  is restricted to qubit or qutrit gates.
Non-unitary  super-quantum  cloning or  deleting,  introduced in  Ref.
\cite{sen}, which  lead to superluminal signaling,  are other examples
of non-complete operations.

Even at the single-particle level, if the measurement is non-complete,
there   is   a   superluminal    signaling   due   to   breakdown   in
non-contextuality  coming  from  the  renormalization.   As  a  simple
illustration, suppose we are given two observers Alice and Bob sharing
a       delocalized       qubit,      $\cos(\theta/2)|0\rangle       +
\sin(\theta/2)|1\rangle$,  with eigenstate $|1\rangle$  localized near
Alice and $|0\rangle$  near Bob.  With an $m$-fold  application of $G$
(which  can be  thought of  as an  application of  imaginary  phase on
Alice's side, leading to  selective augmentation of amplitude) on this
state,      Alice      produces      the     (unnormalized)      state
\mbox{$\cos(\theta/2)|0\rangle      +       (1      +      \epsilon)^m
  \sin(\theta/2)|1\rangle$},  so  that  after  renormalization,  Bob's
probability    of   obtaining   $|0\rangle$    has   changed    in   a
context-dependent fashion from $\cos^2(\theta/2)$ to $\cos^2(\theta/2)
(\cos^2(\theta/2)  +  (1+\epsilon)^{2m}  \sin^2(\theta/2))^{-1}$.   By
thus  nonlocally  controlling the  probability  with  which Bob  finds
$|0\rangle$, Alice  can probabilistically signal 1  bit of information
superluminally.

(b)  {\em  Nonlinear  evolution.}   As  a  simple  illustration  of  a
superluminal gate  arising from nonlinear evolution,   we consider the
action of  the nonlinear  two-qubit `OR' gate  $R$, whose action  in a
preferred (say, computational) basis is given by:
\begin{equation}
\label{eq:nonlinor}
\left.
\begin{array}{ll}
|00\rangle \pm |11\rangle \\
|01\rangle \pm |10\rangle \\
|01\rangle \pm |11\rangle \\
\end{array} \right\} \stackrel{R}{\longrightarrow}
|01\rangle \pm |11\rangle; \hspace{1.0cm}
\begin{array}{l}
|00\rangle \pm |10\rangle 
\stackrel{R}{\longrightarrow} |00\rangle \pm |10\rangle), \\
|\alpha\beta\rangle \stackrel{R}{\longrightarrow} |\alpha\beta\rangle.
\end{array}
\end{equation}
If the  two qubits are entangled  with other qubits, then  the gate is
assumed to act in each subspace labelled by states of the other qubits
in the computational  basis.  Alice and Bob share  the entangled state
$|\Psi\rangle = 2^{-1/2}(|00\rangle -  |11\rangle)$. To transmit a bit
superluminally Alice measures her  qubit in the computational basis or
the  diagonal  basis   $\{|\pm\rangle  \equiv  2^{-1/2}(|0\rangle  \pm
|1\rangle\}$,   leaving   Bob's   qubit's   density  operator   in   a
computational basis  ensemble or a diagonal basis  ensemble, which are
equivalent in standard QM.  However, with the nonlinear operation $R$,
the  two ensembles  can be  distinguished. Bob  prepares  an ancillary
qubit in  the state $|0\rangle$,  and applies a  CNOT on it,  with his
system qubit  as the control. On  the resulting state  he performs the
nonlinear  gate  $R$, and  measures  the  ancilla.  The  computational
(resp., diagonal)  basis ensemble yields the value  1 with probability
$\frac{1}{2}$ (resp.,  1). By  a repetition of  the procedure  a fixed
number $m$ of  times, a superluminal signal is  transmitted from Alice
to Bob with exponentially small  uncertainty in $m$.  Analogous to Eq.
(\ref{eq:nonlinor}), one  can define a `nonlinear  AND', which, again,
similarly leads  to a  nonlocal signaling. Even  at a  single particle
level, allowing for  non-complete operations, superluminal effects can
arise from the nonlinearity due to renormalization \cite{shiekh}.

(c)  {\em  Departure  from  the  Born  $|\psi|^2$  probability  rule.}
Gleason's theorem shows that the Born probability rule that identifies
$|\psi|^2$  as a probability  measure, and  more generally,  the trace
rule, is the  only probability prescription consistent in  3 or larger
dimensions     with    the     requirement     of    non-contextuality
\cite{gle57}. Suppose we retain  unitary evolution, which preserve the
2-norm, but assume that the  probability of a measurement on the state
$\sum_j  \alpha_j|j\rangle$   is  of  the   form  $|\alpha_j|^p/\sum_k
|\alpha_k|^p$ for  outcome $j$, and $p$ any  non-negative real number.
The renormalization will make  the measurement contextual, giving rise
to a  superluminal signal.  One might consider  more general evolution
that preserves a  $p$-norm, but there are no  linear operators that do
so except permutation matrices \cite{insel}.

For example,  let Alice  and Bob share  the two-qubit  entangled state
$\cos\theta|00\rangle + \sin\theta|11\rangle$  ($0 < \theta < \pi/2)$.
The probability for Alice  measuring her particle in the computational
basis and finding $|0\rangle$ (resp., $|1\rangle$) must be the same as
that  for a  joint measurement  in  this basis  to yield  $|00\rangle$
(resp.,  $|11\rangle$). Therefore  Bob's reduced  density  operator is
given by  the state  $\rho^{(1)} = (\cos^p\theta  |0\rangle\langle0| +
\sin^p\theta  |1\rangle\langle1|)/(\cos^p\theta +  \sin^p\theta)$.  On
the other hand, if Alice employs an ancillary, third qubit prepared in
the state $|0\rangle$, and applies a Hadamard on it conditioned on her
qubit  being  in the  state  $|0\rangle$,  she  produces the  state  $
\frac{\cos\theta}{\sqrt{2}}|000\rangle                                +
\frac{\cos\theta}{\sqrt{2}}|001\rangle  + \sin\theta|110\rangle$.  The
probability that Alice obtains outcomes 00,  01 or 10 must be that for
a joint measurement to yield 000,  001 or 110.  Along similar lines as
in the above case we find that she leaves Bob's qubit in the state
\begin{equation}
\rho^{(2)} \equiv
\frac{2^{(1-p/2)}\cos^p\theta     |0\rangle\langle0|     +    \sin^p\theta
|1\rangle\langle1|)}
{2^{(1-p/2)}\cos^p\theta  + \sin^p\theta}.  
\end{equation}
Since    $\rho^{(1)}$   and    $\rho^{(2)}$    are   probabilistically
distinguishable, with sufficiently many shared copies Alice can signal
Bob one bit superluminally, unless $p=2$.

\section{Exponential gates \label{sec:sripol}}

As  superluminal  quantum  gates   like  $G$  or  $R$  are  internally
consistent, one can  consider why no such operation  occurs in Nature,
whether a  fundamental principle prevents  their physical realization.
One     candidate    principle     is    of     course    no-signaling
itself. Alternatively, since we would  like to derive it, linearity of
QM  may  be taken  as  an axiom.   Since  all  the above  non-standard
operations involve  an overall nonlinear evolution,  the assumption of
strict  quantum   mechanical  linearity  can  indeed   rule  out  such
non-standard  gates. Yet  it  must  be admitted  that,  from a  purely
physics  viewpoint, assuming  that  QM is  linear  affords no  greater
insight than assuming it to  be a non-signaling theory.  We would like
to  suggest  that  the the  absence  of  such  operations may  have  a
complexity theoretic basis.

Both  superluminal gates  as  well as  hypothetical  gates that  allow
efficient  solving  of  intractable  problems  involve  some  sort  of
communication  across  superposition  branches.   In  particular,  the
superluminal gates of Section  \ref{sec:srisum} can be turned into the
latter type of gates, as discussed below.

(a) {\em Non-complete quantum gates.}  It is easily seen that the gate
$G$  in  Eq. (\ref{eq:g})  can  be  used  to solve  {\bf  NP}-complete
problems efficiently.  Consider  solving boolean satisfiability (SAT),
which is {\bf NP}-complete:  given an efficiently computable black box
function $f: \{0,1\}^n \mapsto  \{0,1\}$, to determine if there exists
$x$ such  that $f(x) =  1$.  With the  use of an oracle  that computes
$f(x)$, we prepare the $(n+1)$-qubit entangled state
\begin{equation}
|\Psi_{nc}\rangle = 2^{-n/2}\sum_{x\in\{0,1\}^n}|x\rangle|f(x)\rangle, 
\label{eq:sat}
\end{equation}
and then apply  $G^m$ to the second, 1-qubit register,  where $m$ is a
sufficiently  large  integer,   before  measuring  the  register.   In
particular,   suppose  that   at  most   one  solution   exists.   The
un-normalized  `probability  mass'  of obtaining  outcome  $|1\rangle$
becomes  1  (and the  normalized  probability  about  1/2) when  $m  =
n/(2\log(1+\epsilon))$, if there is a solution, or, if no solution
exists,  remains 0.   Repeating the  experiment a  fixed  number of
times, and applying the Chernoff bound,  we find that to solve SAT, we
only require $m  \in O(n)$.  For terminological brevity,  we will call
as  `exponential  gate' such  a  non-standard  gate  that enables  the
efficient computation of {\bf NP}-complete problems, and denote by $E$
the set  of all exponential gates,  restricted in the  present work to
qubits  and qutrit  gates. 

(b) {\em Nonlinear quantum gates.}  The nonlinear operation $R$ in Eq.
(\ref{eq:nonlinor})    can   be    used   to    efficiently   simulate
nondeterminism.     We   prepare    the   state    $|\psi\rangle$   in
Eq. (\ref{eq:sat}), where the first  $n$ qubits are called the `index'
qubits and the  last one the `flag' qubit.   There are $2^{n-1}$ 4-dim
subspaces, consisting  of the  first index qubit  and the  flag qubit,
labelled by the index qubits $2,\cdots,n$.  On each such subspace, the
first index qubit and flag qubit  are in one of the states $|00\rangle
+ |11\rangle$,  $|01\rangle + |10\rangle$,  $|00\rangle + |10\rangle$.
The operation  Eq.  (\ref{eq:nonlinor}) is applied  $n$ times, pairing
each index qubit sequentially with  the flag. The number of terms with
1 on  the flag bit doubles with  each operation so that  after the $n$
operations, it becomes disentangled and can then be read off to obtain
the  answer \cite{bram98}.   A slight  modification of  this algorithm
solves  $\#{\bf P}$-complete  problems efficiently,  by  replacing the
flag qubit with  $\log n$ qubits and the  1-bit nonlinear OR operation
with the corresponding nonlinear  counting.  The final readout is then
the  number  of solutions  to  $f(x)=1$  \cite{bram98}.  Applying  the
nonlinear OR and AND alternatively  to the state $|\psi\rangle$ in Eq.
(\ref{eq:sat}) allows one to  efficiently solve the quantified Boolean
formula  problem,   which  is  {\bf   PSPACE}-complete  \cite{pspace}.
Furthermore, it can  be shown that a single  particle quantum computer
employing  the nonlinear  (due to  renormalization)  quantum mechanism
mentioned  above,  enables  efficient  solution of  {\bf  NP}-complete
problems \cite{shiekh}.

(c)  {\em  Non-Gleasonian  gates.}   By  employing  polynomially  many
ancillas in the method of (c) in the previous subsection, one can show
that non-Gleasonian quantum computers (for  which $p \ne 2$) can solve
{\bf PP}-complete  problems \cite{sripp} efficiently.   Defining ${\bf
BQP}_p$  as similar  to ${\bf  BQP}$, except  that the  probability of
measuring  a  basis   state  $|x\rangle$  equals  $|\alpha_x|^p/\sum_y
|\alpha_y|^p$ (so  that ${\bf  BQP}_2 = {\bf  BQP}$), it can  be shown
that ${\bf PP} \subseteq {\bf BQP}_p$ for all constants $p \ne 2$, and
that, in particular,  ${\bf PP}$ exactly characterizes the  power of a
quantum computer with even-valued $p$ (except $p=2$) \cite{insel}.

In view  of the connection  between the two  classes of gates,  we now
propose, as we earlier did  in Ref.  \cite{sri06}, that the reason for
the  absence   in  Nature  of   the  superluminal  gates   of  Section
\ref{sec:srisum} is  WNHE: in a  universe that is a  polynomial place,
exponential gates like $G$ and $R$ are ruled out.  In the next Section
we  will  consider  in  further  detail  the  viability  of  the  WNHE
assumption as an explanation for no-signaling.

\section{Polynomial superluminal gates \label{sec:sripolynon}}

Even though  WNHE excludes the  type of superluminal  gates considered
above, for the exclusion to be general, it would have to be shown that
every superluminal  gate is exponential, i.e., $C^<  \subseteq E$.  It
turns  out  that  this  cannot  be done,  because  one  can  construct
hypothetical  {\it polynomial superluminal  gates},  which are  superluminal
operations that are not exponential. In fact, it is probably true that
$E  \subset C^<$.   To  see this,  let  us consider  solving the  {\bf
NP}-complete  problem associated with  Eq.  (\ref{eq:sat})  via Grover
search \cite{srigro97},  which is  optimal for QM  \cite{sriben97} but
offers only a  quadratic speed-up, thus leaving the  complexity of the
problem exponential  in $n$, at least  in the black  box setting.  The
optimality  proof  relies  on  showing  that,  given  the  problem  of
distinguishing an  empty oracle  ($\forall_x A(x)=0$) and  a non-empty
oracle containing a  single random unknown string $y$  of known length
$n$ (i.e.   $A(y) = 1$, but  $\forall_{x\ne y} A(x) =  0$), subject to
the constraint that  its overall evolution be unitary,  and linear (so
that in  a computation with  a nonempty oracle, all  computation paths
querying empty  locations evolve  exactly as they  would for  an empty
oracle), the speed-up over a classical search is at best quadratic.

Any degree  of amplitude amplification  of the marked state  above the
quadratic level would then  require empty superposition branches being
`made  aware'  of  the  presence   of  a  non-empty  branch,  i.e.,  a
nonlinearity  of  some  sort.   Let  us  suppose  Bob  can  perform  a
trace-preserving   nonlinear    transformation   $\rho_j   \rightarrow
\tilde{\rho}_j$ of the above kind  on an unknown ensemble of separable
states.  Further, let Alice and Bob share an entangled state, by which
Alice is able to prepare, employing two different POVMs, two different
but equivalent  ensembles of Bob.  Then, depending  on Alice's choice,
his  reduced density  matrix  evolves as  $\rho_B  = \sum_j  p_j\rho_j
\rightarrow \sum  p_j \tilde{\rho}_j \equiv \rho^\prime$  or $\rho_B =
\sum_s   p_k\rho_k   \rightarrow   \sum  p_k   \tilde{\rho}_k   \equiv
\rho^{\prime\prime}$  where   $(\rho_j,p_j)$  and  $(\rho_k,p_k)$  are
distinct,  equivalent ensembles  \cite{sriper02}.   The assumption  of
linearity    is   sufficient   to    ensure   that    $\rho^\prime   =
\rho^{\prime\prime}$.   This  is not  guaranteed  in  the presence  of
nonlinearity,  leading  to  a  potential superluminal  signal.   In  a
nonlinearity of the above kind, the result would depend on whether the
particular ensemble remotely prepared by Alice has states that include
$|y\rangle$  in the  superposition.   This would  lead  to a  scenario
similar  to  that  encountered  with  nonlinear gate  $R$  in  Section
\ref{sec:srisum}.


Possibly the  simplest examples  of polynomial superluminal  gates are
the non-invertible  {\em constant  gates}, which map  any state  in an
input Hilbert space to a fixed  state in the output Hilbert space, and
have the form $|\xi\rangle \otimes  \sum_j \langle j|$, for some fixed
$\xi$. Examples in matrix notation are:
\begin{equation}
Q =  \left(\begin{array}{ll}   1  &  1  \\  0  &  0
\end{array}\right);\hspace{0.5cm}
Q ^{\prime} 
= \left(\begin{array}{lll} 1 & 1 & 1 \\ 0 & 0 & 0 \\ 0 & 0 & 0
\end{array}\right),
\label{eq:Kjanichwara}
\end{equation}
acting      in     Hilbert      space     ${\cal      H}_2     \equiv$
span$\{|0\rangle,|1\rangle\}$     and      ${\cal     H}_3     \equiv$
span$\{|0\rangle,|1\rangle,|2\rangle\}$, respectively.   They have the
effect of  mapping any input state  in ${\cal H}_2$ to  a fixed (apart
from  a  normalization  factor)  state  $|\xi\rangle$,  in  this  case
$|\xi\rangle$ being $|0\rangle$.  In Eq. (\ref{eq:Kjanichwara}), we do
not in general 
require the input and output bases to be the same, nor indeed that
the input  and output Hilbert subspaces  be the same  (for example, as
with  the  distinct  incoming  and  outgoing  modes  of  a  scattering
problem.)

Both $Q$  and $ Q^{\prime}$  are non-complete, inasmuch  as $Q^{\dag}Q
\ne  \mathbb{I}$ and  $(Q^\prime)^\dag Q^\prime  \ne  \mathbb{I}$, and
represent  superluminal  gates.   For  example,  by  applying  or  not
applying $Q$  to her register in the  state $(1/\sqrt{2})(|01\rangle +
|10\rangle)$ shared with Bob, Alice  can remotely prepare his state to
be the pure state $(1/\sqrt{2})(|0\rangle + |1\rangle)$ or leave it as
a maximal  mixture, respectively. Similarly, by choosing  to apply, or
not, $Q^\prime$  on her half  of the state  $(1/\sqrt{2})(|11\rangle +
|22\rangle)$ shared  with Bob, Alice  can superluminally signal  him 1
bit.  

The constant gate is linear and presumes no re-normalization following
its  non-complete  action.  The  probability  of  the  occurance of  a
constant  gate $C$ when  it is  applied to  a state  $|\psi\rangle$ is
simply given by $||C|\psi\rangle||^2$,  per the usual prescription.  One
consequence  is that  it could  not  be used  to violate  no-signaling
without the use of entanglement.  As an illustration: in ${\cal H}_3$,
let the states $|0\rangle$ and $|1\rangle$ be localized near Alice and
$|2\rangle$   near   Bob.    Applying   $Q^{\prime}$  on   the   state
$|\psi\rangle  \equiv  a|0\rangle +  b|1\rangle  + c|2\rangle$,  Alice
obtains  the  (unnormalized)  state  $Q^{\prime}|\psi\rangle  =  (a  +
b)|0\rangle  + c|2\rangle$.   If renormalization  were  allowed, Alice
could nonlocally influence Bob's probability to find $|2\rangle$ to be
$|c|^2/(|a+b|^2+|c|^2)$  or $|c|^2$.   However, the  linearity  of the
constant gate  requires the interpretation that  following her action,
Alice can  detect the particle  with probability $|a+b|^2$,  while for
Bob,  the probability  remains $|c|^2$.  As clarified  later,  lack of
probability conservation can be interpreted as coherent enhancement or
suppression of emission of particles from a source to a detector.

On the other  hand, neither $Q$ nor $Q^\prime$  nor a general constant
gate is an exponential gate:  each of them simply transforms any valid
input into a  fixed output.  Intuitively,
this lack of any  dependence on the input
clearly limits its computational power.   
The  family of  constant gates like $Q$ and $Q^\prime$ 
is simply  equivalent  to a
non-deterministic  simulation  of  a  constant function, and can  be
simulated by  the following {\em classical}  Turing machine pseudocode
in polynomial time (in fact, $O(1)$ time): ``Read first bit of $x$; if
$x \ne$ NULL, output 0; else output NULL".

Operations  $Q$ and  $Q^\prime$ in  Eq. (\ref{eq:Kjanichwara})  can be
extended to a more general  class of polynomial superluminal qubit and
qutrit gates
\begin{equation}
Q_2(\phi) =  \left(\begin{array}{ll}   1  &  e^{i\phi}  \\  0  &  0
\end{array}\right),\hspace{0.5cm}
Q_3(\phi_1,\phi_2)
= \left(\begin{array}{lll} 1 & e^{i\phi_1} & e^{i\phi_2} 
\\ 0 & 0 & 0 \\ 0 & 0 & 0
\end{array}\right), ~{\rm etc.}
\label{eq:Kjanichwara2}
\end{equation}

By  definition,  $Q=Q_2(0)$  and  $Q^\prime=Q_3(0,0)$.   To  see  that
$Q_2(\phi)$ is a polynomial operation, it suffices to show that it can
be  simulated  using  only   polynomial  amount  of  standard  quantum
mechanical   resources.   Given   an  arbitrary   $(n+1)$-qubit  state
$|\psi\rangle  =  |\alpha\rangle|0\rangle  +  |\beta\rangle|1\rangle$,
where  $|\alpha\rangle$   and  $|\beta\rangle$  are   not  necessarily
mutually  orthogonal nor normalized,  one first  applies a  phase gate
$\left( \begin{array}{cc}  1 & 0 \\  0 & e^{i\phi}\end{array}\right)$,
followed by a  Hadamard on the $n$th qubit,  followed by a measurement
conditioned  on  the outcome  being  $|0\rangle$,  which happens  with
probability    $(|||\alpha\rangle||^2    +    |||\beta\rangle||^2    +
\Re[e^{i\phi}\langle\alpha|\beta\rangle])/2$, irrespective of $n$.  If
$|\alpha\rangle$  and $|\beta\rangle$  are orthogonal,  the simulation
succeeds with  fixed probability $\frac{1}{2}$.   Therefore, the class
of  problems efficiently solvable  using quantum  computation equipped
with the non-standard family of constant gates is in {\bf BQP}.

In point of fact, one could be worse off applying a constant gate than
not applying it. In Eq.  (\ref{eq:sat}), let $|\Psi_{\rm nc}^1\rangle$
represent  the  state derived  for  the  function $f^1(\cdot)$,  where
$f^1(j) =1$  for precisely one $j$, and  let $|\Psi_{\rm nc}^0\rangle$
represent  the  state derived  for  the  function $f^0(\cdot)$,  where
$\forall_j f(j)=0$.  In both  cases, upon applying $\mathbb{I} \otimes
Q$,   we  obtain   the  same   disentangled   state,  $2^{-n/2}(\sum_j
|j\rangle)|0\rangle$.  The application of  $Q$ causes the distance and
hence  distinguishability   between  the   two states to diminish,  or
equivalently,  the  fidelity  between them to increase: $1 =  |\langle
\Psi_{nc}^1(\mathbb{I}    \otimes   Q^\dag)(\mathbb{I}    \otimes   Q)
|\Psi^0_{\rm    nc}\rangle|    >   |\langle    \Psi_{nc}^1|\Psi^0_{\rm
nc}\rangle| = 1 - O(2^{-n})$.


It is worth noting that the  constant gate is quite different from the
following two  operations that appear to  be similar, but  are in fact
quite distinct.  The first operation is a  standard quantum mechanical
completely positive  map, polynomial and not  superluminal; the second
is exponential and consequently superluminal.

(a)  To   begin  with, a  constant   gate  is  not  a   quantum  deleter
\cite{sri07},  in which  a  qubit  is subjected  to  a {\em  complete}
operation,  in specific,  a contractive  completely positive  map that
prepares it  asymptotically in a fixed state  $|0\rangle$.  The action
of  a  quantum  deleter  is  given by  an  amplitude  damping  channel
\cite{nc00}, which has an operator sum representation, respectively
\begin{equation}
\label{eq:qdele}
\rho_2 \longrightarrow  \sum_j E_j \rho_2  E_j^{\dag};\hspace{0.5cm}
\rho_3 \longrightarrow \sum_j E_j^{\prime} \rho_3 E_j^{\prime\dag},
\end{equation}
in  the qubit  case or when extended to  the qutrit case, 
with  the Kraus operators  given by Eq.
(\ref{eq:sriKjanichwaraA}) or (\ref{eq:sriKjanichwaraB}), respectively
\begin{subequations}
\label{eq:sriKjanichwara}
\begin{eqnarray}
E_1 &\equiv& \left(\begin{array}{ll} 1 & 0 \\ 0 & 0
\end{array}\right),~~ E_2 \equiv \left(\begin{array}{ll} 0 & 1 \\ 0 & 0 
\end{array}\right), \label{eq:sriKjanichwaraA} \\ E_1^{\prime} &\equiv&
\left(\begin{array}{lll} 1 & 0 & 0 \\ 0 & 0 & 0 \\ 0 & 0 & 0
\end{array}\right),~~ E_2^{\prime} \equiv \left(\begin{array}{lll} 
0 & 1 & 0 \\ 0 & 0 & 0 \\ 0 & 0 & 0 \end{array}\right),~~ E_3^{\prime} 
\equiv \left(\begin{array}{lll} 0 & 0 & 1 \\ 0 & 0 & 0 \\ 0 & 0 & 0 
\end{array}\right).
\label{eq:sriKjanichwaraB}
\end{eqnarray}
\end{subequations}
Unlike in the case of  $Q$, $Q^\prime$ or $Q^{\prime\prime}$, there is
no actual destruction of quantum information, but its transfer through
dissipative  decoherence  into  correlations with  a  zero-temperature
environment.  The  reduced density operator of  Bob's entangled system
remains  unaffected by Alice's  application of  this operation  on her
system.   The deleting action,  though nonlinear  at the  state vector
level, nevertheless acts linearly on the density operator.
  

(b) Next  we note that the  constant gate is quite  different from the
`post-selection'  operation,  which is  a  {\it deterministic}  rank-1
projection \cite{insel}.   Verbally, if the  constant gate corresponds
to  the   operation  ``for  all   input  states  $|j\rangle$   in  the
computational   basis,  set   the  output   state   to  $|\xi\rangle$,
independently of $j$, except  for a global phase", where $|\xi\rangle$
is  some fixed state,  then post-selection  corresponds to  the action
``for  all input  states $|j\rangle$,  if  $j \ne  \xi$, then  discard
branch $|j\rangle$".  Post-selective equivalents of $Q$ and $Q^\prime$
are
\begin{equation}
Q_{\bf PS} =  \left(\begin{array}{ll}   1  &  0  \\  0  &  0
\end{array}\right);\hspace{0.5cm}
Q ^{\prime}_{\rm PS} 
= \left(\begin{array}{lll} 1 & 0 & 0 \\ 0 & 0 & 0 \\ 0 & 0 & 0
\end{array}\right),
\label{eq:KjanichwaraS}
\end{equation}
followed by renormalization.  In particular, whereas the action of $Q$
on the first of two  particles in the state $(1/\sqrt{2})(|00\rangle +
|11\rangle)$    leaves   the    second   particle    in    the   state
$(1/\sqrt{2})(|0\rangle + |1\rangle)$, that of $Q_{\rm PS}$ leaves the
second particle  in the state  $|0\rangle$.  It is  straightforward to
see that post-selection is an  exponential operation: acting it on the
second  qubit  of   $|\Psi_{nc}\rangle$  in  Eq.  (\ref{eq:sat}),  and
post-selecting on 1, we obtain the solution to SAT in one time-step.

The seemingly immediate conclusion  due to the fact $C^< \not\subseteq
E$  is  that  the WNHE  assumption  is  not  strong enough  to  derive
no-signaling,  and  would  have  to be  supplemented  with  additional
assumption(s), possibly purely  physically motivated ones, prohibiting
the physical realization of polynomial superluminal gates.

An alternative, highly unconventional reading of the situation is that
WNHE  is a  fundamental principle  of  the physical  world, while  the
no-signaling  condition  is  in  fact  not  universal,  so  that  some
polynomial   superluminal    gates   may   actually    be   physically
realizable. Quite surprisingly,  we may be able to  offer some support
for this viewpoint.  We believe  that constant gates of above type can
be quantum  optically realized  when a photon  detection is made  at a
{\em path singularity}, defined as a  point in space where two or more
incoming paths converge and terminate.  In graph theoretic parlance, a
path singularity  is a  terminal node in  a directed graph,  of degree
greater than  1.  

We  describe  in Section  \ref{sec:ent}  an  experiment that  possibly
physically realizes $Q$.  In principle, a detector placed at the focus
of a  convex lens realizes such  a path singularity.   This is because
the geometry of the ray  optics associated with the lens requires rays
parallel to the  lens axis to converge to  the focus after refraction,
while  the   destructive  nature  of  photon   detection  implies  the
termination of the path.   As another example, consider a Mach-Zehnder
interferometer  where  the  second  beam-splitter  is  replaced  by  a
detector: the  two converging arms  are then brought into  overlap and
detected in the  overlap region, without being sent  through, as would
be  the case  in a  conventional  Mach-Zehnder set-up.   We find  that
although conceptually  and experimentally  simple, the high  degree of
mode filtering  or spatial  resolution that these  experiments require
will be the  main challenge in implementing them.   Indeed, we believe
this is the reason that such gates have remained undiscovered so far.

Our argument here has implicitly  assumed that ${\bf P} \ne {\bf NP}$.
If it  turns out that  ${\bf P} =  {\bf NP}$, then even  the obviously
non-physical operations such as $G$  or $R$ would be polynomial gates,
and  the  WNHE   assumption  would  not  be  able   to  exclude  them.
Nevertheless,  the question  of existence  and testability  of certain
superluminal gates, which is the main result of this work, would still
remain valid  and of interest.   If polynomial superluminal  gates are
indeed found to exist (and  given that other superluminal gates do not
seem  to exist  anyway), this  would give  us greater  confidence that
${\bf P}  \ne {\bf  NP}$ (or, to  be safe,  that even Nature  does not
`know' that ${\bf P} = {\bf NP}$!)  and that the assumption of WNHE is
indeed a valid and fruitful one.

\section{An experiment with entangled pairs of photons \label{sec:ent}}

\begin{figure}
\includegraphics[width=17.0cm]{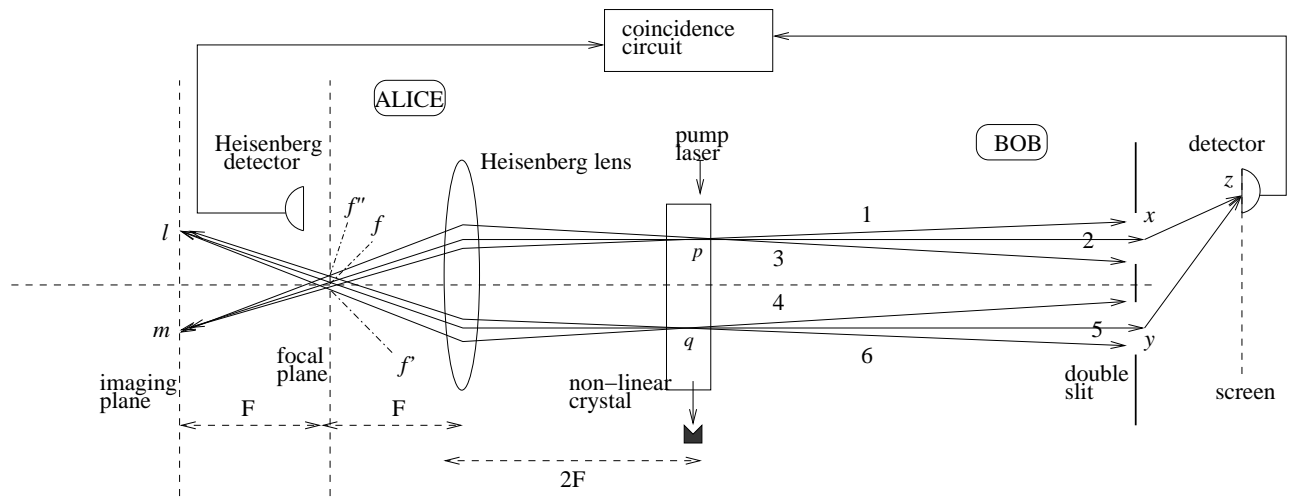} 
\caption{An  `unfolded' version  of the  Innsbruck experiment  (not to
  scale).  A  pair of momentum-entangled photons is  created by type-I
  parametric down  conversion of the pump laser.   Alice's photon (the
  signal  photon) is registered  by a  detector behind  the Heisenberg
  lens.   Bob's photon (the  idler) is  detected behind  a double-slit
  assembly.  If the `Heisenberg detector' is placed in the focal plane
  of the  lens (of focal length  $F$), it projects Bob's  state into a
  mixture  of plane waves,  which produce  an interference  pattern on
  Bob's screen {\em in coincidence}  with any fixed detection point on
  Alice's focal plane. Bob's pattern in his {\em singles count}, being
  the integration of such patterns  over all focal plane points, shows
  no  interference  pattern.   On  the  other  hand,  positioning  the
  Heisenberg detector in the  imaging plane can potentially reveal the
  path the  idler takes through the  slit assembly, and  thus does not
  lead  to  an  interference  pattern  on Bob's  screen  even  in  the
  coincidence counts.}
\label{fig:zeiz}
\end{figure}

Our proposed implemention  of $Q^\prime$, based  on the  use of
entanglement,  is  broadly related  to  the  type  of quantum  optical
experiments encountered in Refs.   \cite{asp8}, and closely related to
an  experiment  performed   in  Innbruck  that  elegantly  illustrates
wave-particle duality by  means of entangled light \cite{zei00,dop98}.
In  the  Innsbruck experiment,  pairs  of position-momentum  entangled
photons  are  produced  by  means  of  type-I  spontaneous  parametric
down-conversion (SPDC) at  a nonlinear source, such as  a BBO crystal.
The two outgoing conical beams from the nonlinear source are presented
`unfolded'  in Figure \ref{fig:zeiz}.   One of  each pair,  called the
`signal photon',  is received  by Alice, while  the other,  called the
`idler',  is  received  and   analyzed  by  Bob.   Alice's  photon  is
registered by a detector behind a `Heisenberg lens'.

Bob's photon is  detected after it enters a  double-slit assembly.  If
Alice's detector, which  is located behind the lens,  is positioned at
the focal plane of the lens and detects a photon, it localizes Alice's
photon to a point on the focal plane.  By virtue of entanglement, this
projects the state of Bob's photon to a `momentum eigenstate', a plane
wave  propagating in a  particular direction.   For example,  if Alice
detects  her photon  at $f$,  $f^\prime$ or  $f^{\prime\prime}$, Bob's
photon is projected to a superposition  of the parallel modes 2 and 5,
modes 1 and 4, or modes  3 and 6.  Since this cannot reveal positional
information about whether  the particle originated at $p$  or $q$, and
hence reveals no which-way  information about slit passage, therefore,
{\em  in coincidence} with  a registration  of her  photon at  a focal
plane  point, the  idler exhibits  a Young's  double-slit interference
pattern  \cite{dop98,zei00}.  The  patterns  corresponding to  Alice's
registering her  photon at $f$, $f^\prime$  or $f^{\prime\prime}$ will
be  mutually shifted.   Bob's observation  in his  single  counts will
therefore not show any sign  of interference, being the average of all
possible such mutually shifted  patterns.  The interference pattern is
seen by Bob in coincidence  with Alice's detection, and cannot be seen
by  him  unilaterally.  This  is  of  course  expected on  account  of
no-signaling.

If the Heisenberg detector is placed at the imaging plane (at distance
$2F$ from the  plane of the lens), a click of  the detector can reveal
the path  the idler takes from  the crystal through  the slit assembly
which  therefore cannot  show  the interference  pattern  even in  the
coincidence counts.  For  example, if Alice detects her  photon at $l$
(resp.,  $m$), Bob's  photon is  projected to  a superposition  of the
mutually  non-parallel modes  4, 5  and  6 (resp.,  1, 2  and 3)  and,
because the  double-slit assembly is  situated in the near  field, can
then  enter only slit  $y$ (resp.,  $x$).  Therefore,  Alice's imaging
plane  measurement gives  path or  position information  of  the idler
photon, so  that no interference pattern emerges  in Bob's coincidence
counts  \cite{dop98,zei00},  and  consequently  also  in  his  singles
counts. This  qualitative description  of the Innsbruck  experiment is
made  quantitative   using  a  simple  six-mode  model   in  the  next
Subsection.

\subsection{Quantum optical description of the Innsbruck experiment 
\label{sec:sridom}}

\begin{figure}
\includegraphics[width=17.0cm]{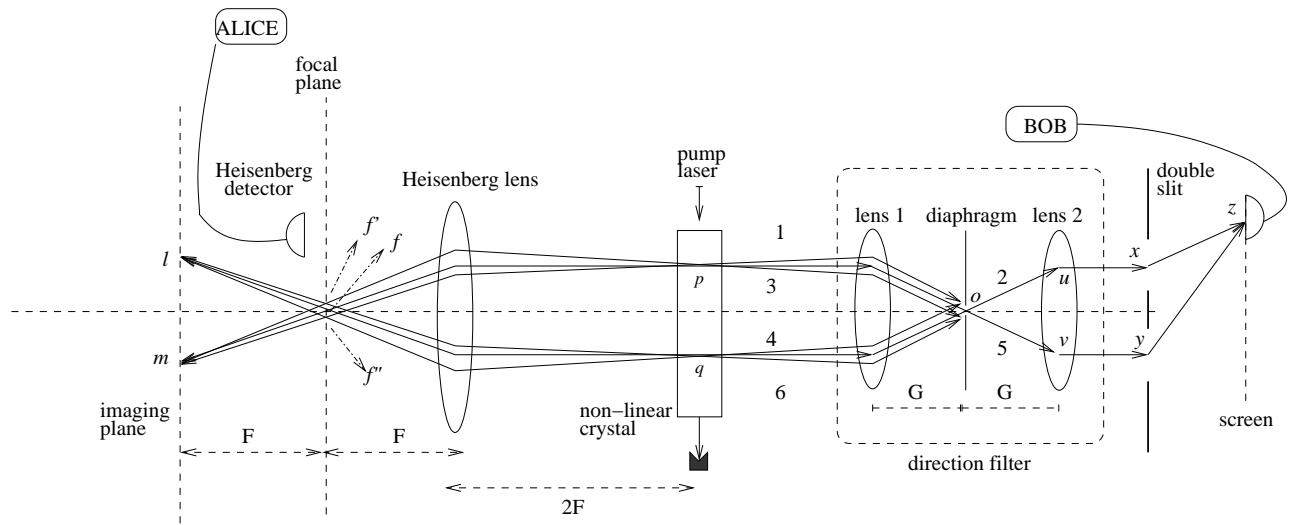} 
\caption{The  modified  Innsbruck  experiment  (not  to  scale):  Same
  configuration as in Figure  \ref{fig:zeiz}, except that Bob's photon
  (the idler),  before entering the double-slit  assembly, traverses a
  direction filter that permits only (nearly) horizontal modes to pass
  through,  absorbing  the  other  modes  at the  filter  walls.   The
  direction  filter acts  as  a  state filter  that  ensures that  Bob
  receives  only the  {\em pure}  state consisting  of  the horizontal
  modes.  Thus  if Alice makes no  measuement or makes  a detection at
  $f$, Bob's  corresponding photon  builds an interference  pattern of
  the modes 2 and 5 in the {\em singles} counts. On the other hand, if
  Alice positions  the Heisenberg detector  in the imaging  plane, she
  knows the  path the idler takes  through the slit  assembly. Thus no
  interference  pattern   is  found  on  Bob's  screen   even  in  the
  coincidence counts.}
\label{fig:zei}
\end{figure}

Here we give a simple, quantitative exposition of the experiment.  The
state of the SPDC field of Figure \ref{fig:zei} is modeled by a 6-mode
vector:
\begin{equation}
\label{spdc}
|\Psi\rangle =  (1 +  \frac{\epsilon}{\sqrt{6}}
\sum_{j=1}^6 a^{\dag}_jb^{\dag}_j)|{\rm vac}\rangle 
\end{equation}
where $|{\rm  vac}\rangle$ is  the vacuum state,  $a^{\dag}_j$ (resp.,
$b^{\dag}_j$) are  the creation  operators for Alice's  (resp., Bob's)
light  field on  mode $j$,  per the  mode numbering  scheme  in Figure
\ref{fig:zei}. The  quantity  $\epsilon$ ($\ll  1$)
depends on the pump field  strength and the crystal nonlinearity.  The
coincidence counting rate for  corresponding measurements by Alice and
Bob  is  proportional  to the  square  of  the  second-order  correlation
function, and given by:
\begin{equation}
\label{Eq:coinc}
R_\alpha(z)      \propto      \langle\Psi|E^{(-)}_z     E^{(-)}_\alpha
E^{(+)}_\alpha  E^{(+)}_z|\Psi\rangle = ||E^{(+)}_\alpha
E^{(+)}_z|\Psi\rangle||^2,   ~~~~   (\alpha   =   f,f^{\prime\prime},l,
m,\cdots).
\end{equation}
where $E_\alpha^{(+)}$  represents the positive frequency  part of the
electric  field at  a point  on Alice's  focal or  imaging  plane, and
$E_z^{(+)}$ that  of the electric field  at an arbitrary  point $z$ on
Bob's  screen.  We have:
\begin{equation}
\label{bobfjeld}
E_z^{(+)} = 
e^{ikr_D}\left(e^{ikr_2}\hat{b}_2 + e^{ikr_5}\hat{b}_5\right) +
e^{ikr_{D^\prime}}\left(e^{ikr_1}\hat{b}_1 + e^{ikr_4}\hat{b}_4\right) +
e^{ikr_{D^{\prime\prime}}}\left(e^{ikr_3}\hat{b}_3 + e^{ikr_6}\hat{b}_6\right), 
\end{equation}
where $k$ is the wavenumber, $r_D$ the distance from the EPR source to
the upper/lower slit on Bob's double slit diaphragm (the length of the
segment $\overline{qy}$  or $\overline{px}$); $r_2$  (resp., $r_5$) is
the distance from the lower (resp., upper) slit to $z$.  The other two
terms in  Eq. (\ref{bobfjeld}),  pertaining to the  other two  pair of
modes,   are  obtained   analogously.    We  study   the  two   cases,
corresponding  to Alice making  a remote  position or  remote momentum
measurement on the idler photons.

{\em Case  1. Alice remotely  measures position (path) of  the idler.}
Suppose Alice positions her dectector at the imaging plane
and detects a photon at $l$ or $m$.
The corresponding field at her detector is
\begin{equation}
\label{alicefjeldb}
E_m^{(+)} =  e^{ik s_{m}}(\hat{a}_1+  \hat{a}_2  + \hat{a}_3); \hspace{0.5cm}  
E_l^{(+)} = e^{ik s_{l}}(\hat{a}_4 + \hat{a}_5 + \hat{a}_6),
\end{equation}
where $s_{m}$ (resp.,  $s_{l}$) is the path length  along any ray path
from the  source point  $p$ (resp., $q$)  through the lens  upto image
point $m$ (resp., $l$). By  Fermat's principle, all paths connecting a
given pair  of source  and image point  are equal.  Setting  $\alpha =
l,m$  in Eq.  (\ref{Eq:coinc}),  and substituting  Eqs.  (\ref{spdc}),
(\ref{bobfjeld}) and (\ref{alicefjeldb})  in Eq.  (\ref{Eq:coinc}), we
find the coincidence counting rate  for detections by Alice and Bob to
be
\begin{equation}
\label{Rg}
R_{m}(z)  \propto   \epsilon^2|e^{ik r_1} + e^{ik r_2} + e^{ik r_3}|^2;
\hspace{0.5cm}
R_{l}(z)  \propto   \epsilon^2|e^{ik r_4} + e^{ik r_5} + e^{ik r_6}|^2,
\end{equation}  
which is essentially a  single slit diffraction pattern formed behind,
respectively,  the upper and lower  slit.  The  intensity  pattern Bob
finds  on his  screen  in  the singles  count,  obtained by  averaging
$R_{\alpha}(z)$   over  $\alpha=l,m$,  is   thus  not   a  double-slit
interference pattern, but an incoherent mixture of the two single slit
patterns.  A similar  lack of interference pattern is  obtained by Bob
if Alice  makes no measurement. 

{\em  Case 2.   Alice remotely  measures momentum  (direction)  of the
  idler.}  Alice  positions her  dectector on the  focal plane  of the
Heisenberg  lens.  If  she  detects  a photon  at  $f$, $f^\prime$  or
$f^{\prime\prime}$, the field at her detector is, respectively,
\begin{subequations}
\label{alicefjelda}
\begin{eqnarray}
E_f^{(+)}  &=&  e^{ikr_{2f}}\hat{a}_2  +  e^{ikr_{5f}}\hat{a}_5
= e^{ikr_{f}}(\hat{a}_2  +  \hat{a}_5),  \label{alicefjeldaa} \\
E_{f^\prime}^{(+)}  &=&  e^{ikr_{1f^\prime}}\hat{a}_1  +  e^{ikr_{4f^\prime}}\hat{a}_4
= e^{ikr_{1f^\prime}}(\hat{a}_1  +  e^{ik(r_{5f^\prime}-r_{1f^\prime})}\hat{a}_4),
\label{alicefjeldab} \\
E_{f^{\prime\prime}}^{(+)}  &=& e^{ikr_{3f^{\prime\prime}}}\hat{a}_3 +
e^{ikr_{6f^{\prime\prime}}}\hat{a}_6                                  =
e^{ikr_{3f^{\prime\prime}}}(\hat{a}_3                                 +
e^{ik(r_{6f^{\prime\prime}}-r_{3f^{\prime\prime}})}\hat{a}_6),
\label{alicefjeldac} 
\end{eqnarray}  
\end{subequations}
where $r_{2f}$ (resp., $r_{5f}$) is the distance from $p$ (resp., $q$)
along the path 2 (resp., 5) path through the lens upto point $f$.  The
distances along the two paths being identical, $r_{2f} = r_{5f} \equiv
r_f$.       The      distances     $r_{1f^\prime},      r_{4f^\prime},
r_{3f^{\prime\prime}}$   and   $r_{6f^{\prime\prime}}$   are   defined
analogously.   Substituting Eqs.   (\ref{spdc}),  (\ref{bobfjeld}) and
(\ref{alicefjelda}) in Eq.   (\ref{Eq:coinc}), we find the coincidence
counting rate is given by
\begin{subequations}
\label{ram}
\begin{eqnarray}
R_f(z)  &\propto& \epsilon^2\left[1 +  \cos(k\cdot[r_2 -  r_5])\right],
\label{rama} \\
R_{f^\prime}(z) &\propto& \epsilon^2\left[1 + \cos(k\cdot[r_1 - r_4] +
  \omega_{14})\right],
\label{ramb} \\
R_{f^{\prime\prime}}(z) &\propto&  \epsilon^2\left[1 + \cos(k\cdot[r_3
    - r_6] + \omega_{36})\right],
\label{ramc}
\end{eqnarray}    
\end{subequations}
where $\omega_{14} \equiv k(r_{4f}  - r_{1f})$ and $\omega_{36} \equiv
k(r_{6f}  -  r_{3f})$  are  fixed  for  a given  point  on  the  focal
plane.  Each equation  in  Eq. (\ref{ram})  represents a  conventional
Young's double slit pattern. Conditioned on Alice detecting photons at
$f$,  Bob  finds  the  pattern  $R_f(z)$,  and  similarly  for  points
$f^\prime$  and   $f^{\prime\prime}$.   In  his   singles  count,  Bob
perceives  no interference,  because  he is  left  with a  statistical
mixture  of  the  patterns (\ref{rama}),  (\ref{ramb}),  (\ref{ramc}),
etc.,  corresponding  to  {\em  all}  points on  Alice's  focal  plane
illuminated by the signal beam.

In summary,  the set-up of  the Innsbruck experiment entails  that Bob
does not find a double-slit  interference pattern in his singles count
no  matter what  Alice does.   However, in  the coincidence  counts he
finds  the interference pattern  if Alice  measures (momentum)  in the
focal plane, and none if she measures (position) in her imaging plane.

\subsection{The proposed experiment \label{sec:sriprop}}

The  experiment  proposed  here,  presented  earlier  by  us  in  Ref.
\cite{sri}, is  derived from  the Innsbruck experiment,  and therefore
called  `the  Modified  Innsbruck  experiment'.   It  was  claimed  to
manifest  superluminal signaling,  though it  was not  clear  what the
exact origin of the signaling was, and in particular, which assumption
that goes to proving the no-signaling theorem was being given up.  The
Modified Innsbruck  experiment is revisited  here in order  to clarify
this issue in  detail in the light of the  discussions of the previous
Sections.   This will  help  crystallize  what is,  and  what is  not,
responsible for the claimed signaling effect.  In Ref.  \cite{sri888},
we  studied  a  version  of  nonlocal communication  inspired  by  the
original   Einstein-Podolsky-Rosen   thought  experiment   \cite{epr}.
Recently, similar experiments, also based on the Innsbruck experiment,
have been independently proposed in Refs. \cite{sricram,sridom}.

First  we present  a qualitative  overview of  the  modified Innsbruck
experiment.   The  only   material  difference  between  the  original
Innsbruck experiment and the modified  version we propose here is that
the  latter contains a  `direction filter',  consisting of  two convex
lenses  of the  same focal  length  $G$, separated  by distance  $2G$.
Their shared focal plane is covered  by an opaque screen, with a small
aperture  $o$ of  diameter $\delta$  at their  shared focus.   We want
$\delta$ to be  small enough so that only  almost horizontal modes are
permitted by  the filter  to fall on  the double slit  diaphragm.  The
angular spread  (about the horizontal) of  the modes that  fall on the
aperture  is given  by $\Delta  \theta  = \delta/G$,  we require  that
$(\delta/G)  \sigma   \ll  \lambda$,   where  $\sigma$  is   the  slit
separation, to guarantee that only modes that are horizontal or almost
horizontal  are selected  to  pass through  the  direction filter,  to
produce  a  Young's double-slit  interference  pattern  on his  screen
plane.  On the  other hand, we don't want the aperture  to be so small
that it produces significant  diffraction, thus: $\delta \gg \lambda$.
Putting these conditions together, we must have
\begin{equation}
1 \ll \frac{\delta}{\lambda} \ll \frac{G}{\sigma}. \label{eq:sricond}
\end{equation}
The  ability  to satisfy  this  condition,  while  preferable, is  not
crucial.  If  it is  not satisfied strictly,  the predicted  signal is
weaker  but not entirely  suppressed. The  point is  clarified further
down.

If Alice  makes no measurement,  the idler remains entangled  with the
signal photon,  which renders incoherent the beams  coming through the
upper  and  lower  slits on  Bob's  side,  so  that  he will  find  no
interference  pattern on his  screen.  Similarly,  if she  detects her
photon in  the imaging plane, she  localizes Bob's photon  at his slit
plane, and so,  again, no interference pattern is  seen. Thus far, the
proposed experiment the same effect as the Innsbruck experiment.

On  the  other hand,  if  Alice  scans the  focal  plane  and makes  a
detection, she remotely measures Bob's corresponding photon's momentum
and erases its path  information, thereby (non-selectively) leaving it
as a mixture of plane waves incident on the direction filter.  However
only  the  fraction  that  makes  up the  pure  state  comprising  the
horizontal modes  passes through  the filter. Diffracting  through the
double-slit diaphragm, it produces  a Young's double slit interference
pattern on  Bob's screen.  Those  plane waves coincident  with Alice's
detecting her photon  away from focus $f$ are filtered  out and do not
reach  Bob's double slit  assembly.  It  follows that  an interference
pattern  will emerge in  Bob's {\em  singles counts},  coinciding with
Alice's detection  at $f$  or close to  $f$.  Thus Alice  can remotely
prepare inequivalent ensembles of  idlers, depending on whether or not
she measures momentum on her photon.  In principle, this constitutes a
superluminal signal.

Quantitatively,  the only  difference  between the  Innsbruck and  the
proposed experiment  is that Eq.  (\ref{bobfjeld}) is replaced  by an
expression containing  only horizontal modes.  As  an idealization (to
be relaxed below), assuming perfect filtering and low spreading of the
wavepacket at the aperature, we have:
\begin{equation}
\label{bobfjeld0}
E_z^{(+)} = 
e^{ikr_D}\left(e^{ikr_2}\hat{b}_2 + e^{ikr_5}\hat{b}_5\right),
\end{equation}
where $r_D$ now represents the distance from the EPR source to
the upper/lower slit on Bob's double slit diaphragm (the length of the
segment $\overline{qoux}$  or $\overline{povy}$); $r_2$  (resp., $r_5$) is
the distance from the upper (resp., lower) slit to $z$.  The other two
{\em Detection of a signal photon  at or near $f$ is the only possible
  event on  the focal plane such  that Bob detects the  twin photon at
  all}.   Focal plane  detections sufficiently  distant from  $f$ will
project the idler into non-horizontal  modes that will be filtered out
before   reaching   Bob's   double-slit  assembly.    Therefore,   the
interference pattern Eq.  (\ref{rama}) is in fact the only one seen in
Bob's singles counts.  We denote  by $R^F(z)$, this pattern, which Bob
obtains  conditioned  on  Alice  measuring  in the  focal  plane.   By
contrast, in the Innsbruck experiment Bob in his singles counts sees a
statistical  mixture  of   the  patterns  (\ref{rama}),  (\ref{ramb}),
(\ref{ramc}), etc., corresponding to {\em all} points on Alice's focal
plane illuminated by the signal beam.

When  Alice  measures  in  the  imaging plane,  as  in  the  Innsbruck
experiment  Bob   finds  no   interference  pattern  in   his  singles
counts.  Setting   $\alpha  =  l,m$  in   Eq.   (\ref{Eq:coinc}),  and
substituting     Eqs.      (\ref{spdc}),     (\ref{bobfjeld0})     and
(\ref{alicefjeldb}) in Eq.   (\ref{Eq:coinc}), we find the coincidence
counting rate for detections by Alice and Bob to be
\begin{equation}
\label{Rg0}
R_{\alpha}(z)  \propto   \epsilon^2,~~~~  (\alpha  =  l,   m),
\end{equation}  
which is a uniform pattern (apart  from an envelope due to single slit
diffraction, which we  ignore for the sake of  simplicity). It follows
that Bob's observed pattern in the singles counts conditioned on Alice
measuring  in the  imaging plane,  $R^I(z)$, is  also the  same, i.e.,
$R^I(z) \propto \epsilon^2$.

Our main  result is the  difference between the patterns  $R^I(z)$ and
$R^F(z)$,  which  implies  that  Alice  can  signal  Bob  one  bit  of
information across the spacelike interval connecting their measurement
events, by choosing to measure her photon in the focal plane or not to
measure.  In practice, Bob  would need to include additional detectors
to  sample or  scan the  $z$-plane  fast enough.   This procedure  can
potentially  form  the basis  for  a  superluminal quantum  telegraph,
bringing into sharp focus  the tension between quantum nonlocality and
special relativity.

Considering the far-reaching implications  of a positive result to the
experiment, we  may pause to consider  whether our analysis  of so far
can be correct, and-- in the chance (however limited) that it is-- how
such a signal may ever arise, in view of the no-signaling theorem.  It
may be easy to dismiss  a proof of putative superluminal communication
as `not even  wrong', yet less easy to spot  where the purported proof
fails and to provide a mechanism for thwarting the signaling. 
For  one, the prediction  of the nonlocal signaling  is based on a
model that  departs only slightly  from our quantum optical  model of
Section \ref{sec:sridom}, which  explains the original Innsbruck 
experiment quite
well. There have been various attempts at proving that quantum
nonlocality  somehow contravenes special  relativity.  The  author has
read some of their accounts, and it was not difficult to spot a hidden
erroneous assumption that led to the alleged conflict with relativity.
Armed with  this lesson,  the present claim  will be different  in the
following three ways:

\begin{itemize}
\item {\em We individually  discuss, in the following Section, various
  possible objections to  our claim, and demonstrate why  each of them
  fails.} By ruling  out all the obvious mechanisms  for thwarting the
  signaling, we are led to believe either (a) that there are erroneous
  but less obvious assumptions that have somehow gone into arriving at
  the superluminal signaling  (more likely), or (b) that  there is new
  physics, associated with the signaling (less likely).

Either way, it  is in the spirit  of science that we must  now rely on
experiments to be the final arbiter  on the question. If item (a) turns out
to be the right scenario  eventually, our present exercise could still
be instructive  in yielding new  theoretical insights. For  example, a
proposal for  superluminal communication based  on light amplification
was eventually  understood to fail because it  violates the no-cloning
theorem, a principle  that had not been discovered at  the time of the
proposal was made (cf. \cite{peres}).

\item {\em We single out, in the following Section, the key assumption
  responsible for  the superluminality.} This  is shown to  be Alice's
  momentum measurement, which implements a non-complete measurement of
  the polynomial superluminal type.   This makes clear exactly what is
  the  non-standard element  at stake, and further makes  it easier  for the
  reader to judge whether the  proposal is wrong, not even wrong, or--
  as we believe is the case-- worth testing experimentally.

\item  {\em We have  furnished computation-  and information-theoretic
  grounds  for why  superluminal gates  could be  possible.}   We have
  shown  how no-signaling  could  be a  nearly-universal-but-not-quite
  side  effect of  the  computation theoretic  properties of  physical
  reality; elsewhere \cite{ganeshchandru},  we show how the relativity
  principle    could   be   a    consequence   of    conservation   of
  information. These  ideas suggest that no-signaling is  not an exact
  or fundamental law, but an  indication of a deeper computational and
  informational layer underlying physical reality.

\end{itemize}

It  would no doubt  be surprising  if such  non-complete measurements,
which have no place in  standard quantum mechanics, turn out to exist.
In the  last Section, we clarify  how they could possibly  fit in with
known  physics.  There  we will  argue that  they arise  owing  to the
potential fact  that practically measurable  quantities resulting from
quantum  field theory  are not  described by  hermitian  operators, at
variance with a key axiom of orthodox quantum theory \cite{srihrv}.

\section{The question of existence and
origin of the signaling \label{sec:sriung}}

In the  Section, we will consider  a number of  possible objections to
our  main result.   It might  at first  be supposed  that as  the only
difference  between  our  set-up   in  Figure  \ref{fig:zei}  and  the
Innsbruck  experiment (Figure  \ref{fig:zeiz}),  the direction  filter
must be responsible for the  signaling, and that therefore, there must
be some  unphysical assumption in the  way the filter  is described to
work.  For example, it might  be supposed that in a legitimate filter,
the spreading  caused by the  aperture would wash out  any information
about  Alice's  choice. Yet,  in  the  case of each objection,  
we will  quantitatively demonstrate why there arises no physical mechanism
to thwart the nonlocal signaling, and thus the  objection fails.  
For instance, contrary to the above example claim,   we will find that
the mode-selection  at the filter can  be described as  a local linear
unitary (and hence  complete) operation acting on the  idler, and thus
should  not lead  to  any violation  of  no-signaling.  The  signaling
arises  from  some  action  of  Alice,  which  we  identify  with  her
`momentum'  measurement,  and which  we  show  to  realize  a  noncomplete
operation in the  subspace of interest.  It turns  out that the filter
only serves the practical purpose of exposing the signaling that would
otherwise remain  hidden in the averaged 
pattern that Bob receives.   These points
are discussed in the following Subsections.

\subsection{Effect of spreading at the direction filter 
\label{sec:ganesh_spread}}

In  an actual  experiment, the  conditions (\ref{eq:sricond})  may not
hold  strictly,  with  narrow   filtering  leading  to  a  diffractive
spreading of Bob's photon.  It might appear that because the direction
filter  localizes the  photon  in  momentum space,  it  would cause  a
complementary positional  spread of the  wavefunction, as a  result of
which  Bob should  observe  a fixed  interference  pattern always,  no
matter what Alice  does (or does not).  However,  a closer examination
shows  that  such   a  spreading  only  causes  a   reduction  in  the
visibility-- and not a total  washout-- of the pattern received by Bob
in the case  of Alice's focal plane measurement.   Thus, the spreading
only lowers-- but does  not eliminate-- the distinguishability between
the two  kinds of pattern  that Bob receives.  A  simple, quantitative
explanation of  this situation is  discussed in the remaining  part of
this Subsection.

For illustration,  suppose we  choose $\delta =  10\lambda$, and  as a
result, nearly only horizontal modes $r_2$ and $r_5$ are selected, but
the diffraction  is strong.   We model this  diffraction as  a unitary
rotation      $\left(\begin{array}{cc}\cos\theta      &     \sin\theta
  \\ -\sin\theta & \cos\theta\end{array}\right)$ in the space acted on
  by $\hat{b}_2$ and $\hat{b}_5$,  where $\theta$ is determined by the
  geometry of  the filter.   In place of  Eq. (\ref{bobfjeld})  we now
  have:
\begin{equation}
\label{eq:Bobfjeldnew}
E_z^{\prime  (+)}  =  e^{ikr_D} \left(e^{ikr_2}(\cos\theta\hat{b}_2  +
\sin\theta\hat{b}_5)  +  e^{ikr_5}(\cos\theta  \hat{b}_5 -  \sin\theta
\hat{b}_2) \right).
\end{equation}
In case of Alice's position measurement, we now have in place of
Eq. (\ref{Rg})
\begin{equation}
\label{SriRgx}
R_\alpha^{\prime}(z)        \propto        \epsilon^2\left[1        \pm
\sin(2\theta)\cos(k\cdot[r_2 - r_5])\right], ~~~
({\rm with~}\pm {\rm ~according~as~} \alpha=l,m),
\end{equation}
where the interference  has arisen because the diffraction  at $o$ has
given rise to an amplitude input  to both slits.  The pattern found by
Bob  in  his singles  counts  is  
\begin{equation}
\label{SriRg}
R_l^{\prime}(z)  + R_m^{\prime}(z) \propto \epsilon^2,
\end{equation}
which is  a constant  pattern (ignoring  the finite
width of the slits), just as when the spreading had
been ignored (Eq. 
(\ref{Rg0})). On the  other hand, in place of Eq. (\ref{ram}), we now obtain
\begin{equation}
\label{sriram}
R^{\prime}_f(z)   \propto   
\epsilon^2\left[1  +   \cos(2\theta)\cos(k\cdot[r_2   -  r_5])\right],
\end{equation}    
We recover  the case of clearest  distinction by setting  $\theta = 0$
(which corresponds to the zero diffraction limit), but even otherwise,
the  two  cases  (\ref{SriRg})  and (\ref{sriram})  are  in  principle
distinguishable   in  terms   of  visibility   (except  in   the  case
$\theta=\pi/4$,  which is  highly unlikely,  and in  any case,  can be
precluded by altering $\delta$ or $G$).

\subsection{Alice's focal plane measurement implements a constant gate} 

The  state (\ref{spdc})  is now  represented in  a simple  way  as the
unnormalized           state           
\begin{equation}
|\Psi^{(1)}\rangle           =
\frac{\epsilon}{\sqrt{6}}\sum_{j=1}^6|j,j\rangle,
\label{eq:psi1}
\end{equation}
where for  simplicity the vacuum  state, which does not  contribute to
the entanglement related  effects, is omitted, and it  is assumed that
each mode  contains at  most one pair  of entangled photons  (i.e., no
higher  excitations  of the  light  field).   Further  because of  the
direction filter, it suffices to restrict our attention to the state
\begin{equation}
|\psi^{(2)}\rangle \propto \frac{1}{\sqrt{2}}(|2,2\rangle  + |5,5\rangle),
\label{eq:srispdc}
\end{equation}
the projection of $|\Psi^{(1)}\rangle$  onto ${\cal H}_2 \otimes {\cal
  H}_2$, where  ${\cal H}_2$ is the subspace  spanned by $\{|2\rangle,
|5\rangle\}$.  Under  these assumptions, Alice's  position measurement
in  this  subspace,  represented  by  the  operators  $\hat{a}_2$  and
$\hat{a}_5$, can  be written as the Kraus  operators $\hat{a}_2 \equiv
|0\rangle\langle  2|$  and $\hat{a}_5  \equiv  |0\rangle \langle  5|$.
Within  ${\cal  H}_{2}$ these  operators  form  a  complete set  since
$\hat{a}^{\dag}_2\hat{a}_2      +      \hat{a}^{\dag}_5\hat{a}_5     =
|2\rangle\langle  2|  + |5\rangle\langle  5|  = \mathbb{I}_2$.   Thus,
Alice's measurement on $|\Psi^{(2)}\rangle$ in the position basis does
not nonlocally affect Bob's reduced density operator in this subspace,
which is proportional to $\mathbb{I}_2/2$.

On  the other  hand, if  Alice  measures momentum,  her measurment  is
represented    by   the    field   operator    $E^{(+)}_f$    in   Eq.
(\ref{alicefjelda}). We have in the above notation
\begin{equation}
E_f^{(+)} \propto \hat{a}_2 + \hat{a}_5
\equiv |0\rangle(\langle 2| + \langle 5|).
\label{eq:srif0}
\end{equation}
This is  just the polynomial  superluminal gate $Q$  in Eq.
(\ref{eq:sriKjanichwara}), with the output basis given by
$\{|0\rangle, |0^\perp\rangle\}$, where $|0^\perp\rangle$ is
any basis element orthogonal to the vacuum state.

By  contrast,  Bob's  measurement,  which involves  no  focussing,  is
complete  (which  rules out  a  Bob-to-Alice superluminal  signaling).
Each  element  of  Bob's  screen  $z$-basis  is  a  possible  outcome,
described  by  the annihilation  operator  approximately  of the  form
$\hat{E}^{(-)}_z  \propto  \hat{a}_2  +  e^{i\gamma}\hat{a}_5$,  where
$\gamma = \gamma(k,z)$ is the phase difference between the paths 2 and
5 from  the slits to a point  $z$ on Bob's screen.   This represents a
POVM  of  the  form  $\hat{E}^{(-)}_z\hat{E}^{(+)}_z  =  (|2\rangle  +
e^{-i\gamma}|5\rangle)  (\langle 2|  + e^{i\gamma}\langle  5|)$.  Even
though  $\hat{E}^{(+)}_z$  has  the  same  form  as  Alice's  operator
$\hat{E}^{(+)}_f$-- as  a Kraus operator describing  the absorption of
two interfering modes at a  point $z$--, yet, when integrated over his
whole `position basis',  Bob's measurement is seen to  form a complete
set,   for,   as   it   can  be   shown,   $\int_{z=-\infty}^{+\infty}
\hat{E}^{(-)}_z\hat{E}^{(+)}_z    dz   =    |2\rangle\langle    2|   +
|5\rangle\langle  5|$. In  the case  of Alice's  momentum measurement,
because  the  detection  happens  at  a path  singularity,  a  similar
elimination of cross-terms via integration is not possible, whence the
non-completeness.  It is indeed somewhat intriguing how geometry plays
a  fundamental  role  in  determining  the completeness  status  of  a
measurement. This has to do with the fact that the direct detection of
a   photon  is   practically   a  determination   of  {\em   position}
distribution.   For example,  even in  remotely measuring  the idler's
momentum, Alice measures her photon's position at the focal plane.  We
will return again to this issue in the final Section.


\subsection{Role of the direction filter \label{sec:sriph}}

A simple model of the action of the perfect direction filter is
\begin{equation}
D  \equiv  \sum_{j=2,5} |j\rangle\langle  j|  + \sum_{j\ne  2,5}
|{\rm-}j\rangle\langle j|
\end{equation}
acting   locally   on   the   second   register  of   the   state   of
Eq. (\ref{eq:psi1}).  Here  $|{\rm -}j\rangle$ can be thought  of as a
state orthogonal to all  $|j\rangle$'s and other $|{\rm -}j\rangle$'s,
that  removes  the  photon   from  the  experiment,  for  example,  by
reflecting it out or by absorption  at the filter. It suffices for our
purpose  to  note that  $D$  can be  described  as  a local,  standard
(linear, unitary and hence complete) operation. Since the structure of
QM  guarantees  that  such   an  operation  cannot  lead  to  nonlocal
signaling,  the conclusion  is  that the  superluminal  signal, if  it
exists, must remain {\em even if the the direction filter is absent}.

We     will    employ     the    notation     $|j+k+m\rangle    \equiv
(1/\sqrt{3})(|j\rangle  + |k\rangle  + |m\rangle)$.   To see  that the
nonlocal  signaling  is implicit  in  the  state  modified by  Alice's
actions  even without  the  application  of the  filter,  we note  the
following:    if   Alice    measures   `momentum'    on    the   state
$|\Psi^{(1)}\rangle$ and detects a  signal photon at $f$, she projects
the corresponding idler into  the state $|2+5\rangle$.  Similarly, her
detection of  a photon at  $f^{\prime\prime}$ projects the  idler into
the state  $|3+6\rangle$, and her detection  at $f^{\prime}$, projects
the idler into the state  $|1+4\rangle$.  Therefore, in the absence of
the  direction  filter,  Alice's  remote measurement  of  the  idler's
momentum  leaves  the idler  in  a  (assumed  uniform for  simplicity)
mixture given by
\begin{equation}
\rho_P \propto |2+5\rangle\langle2+5| + |1+4\rangle\langle1+4|
+ |3+6\rangle\langle3+6|.
\label{eq:srirhop}
\end{equation}
Her momentum measurement is non-complete, since the summation over the
corresponding projectors (r.h.s of  Eq. (\ref{eq:srirhop})) is not the
identity  operation  $\mathbb{I}_6$ pertaining  to  the Hilbert  space
spanned by six modes $|j\rangle$ $(j=1,\cdots,6)$.

On the  other hand, if  Alice remotely measures the  idler's position,
she leaves the idler in the mixture
\begin{equation}
\rho_Q \propto |1+2+3\rangle\langle1+2+3| + |4+5+6\rangle\langle4+5+6|.
\label{eq:srirhoq}
\end{equation}
Here again, her position measurement is non-complete, reflected in the
fact that  the summation over  the corresponding projectors  (r.h.s of
Eq. (\ref{eq:srirhoq})) is not $\mathbb{I}_6$ \cite{sridir}.

Since $\rho_P \ne  \rho_Q$, we are led to  conclude that the violation
of no-signaling {\em is already implicit in the Innsbruck experiment}.
Yet, since Bob measures in the $z$-basis rather than the `mode' basis,
in the absence of a direction filter-- as is the case in the Innsbruck
experiment--,  Bob's screen  will  not register  any  signal, for  the
following  reason.  In case  of Alice's  focal plane  measurement, the
integrated diffraction-interference pattern corresponding to different
outcomes will  wash out any  observable interference pattern.   On the
other hand, in  case of Alice's imaging plane  measurement, Bob's each
detection comes  from the photon's  incoherent passage through  one or
the  other  slit,  and  hence--  again-- no  interference  pattern  is
produced on his  screen.  Thus, measurement at Bob's  screen plane $z$
without   a  direction   filter  will   render   $\rho_P$  effectively
indistinguishable  from $\rho_Q$.   The role  played by  the direction
filter  is to  prevent modal  averaging  in case  of Alice's  momentum
measurement,  by selecting  one set  of  modes.  The  filter does  not
create, but only exposes, a superluminal effect that otherwise remains
hidden.


\subsection{Complementarity of single- and two-particle correlations
\label{sec:corr}}

It  is well  known  that  path information  (or  particle nature)  and
interference (or wave nature) are mutually exclusive or complementary.
In the two-photon case, this  takes the form of mutual incompatibility
of single-  and two-particle interference  \cite{abu01,bos02}, because
entanglement  can be  used to  monitor  path information  of the  twin
particle, and is  thus equivalent to `particle nature'.   One may thus
consider single-  and two-particle correlations as being  related by a
kind   of   complementarity   relation   that  parallels   wave-   and
particle-nature complementarity.  A brief exposition of this idea
is given in the following paragraph.

For a particle in a  double-slit experiment, we restrict our attention
to the Hilbert space ${\cal  H}$, spanned by the state $|0\rangle$ and
$|1\rangle$ corresponding to the upper and lower slit of a double slit
experiment.  Given density operator $\rho$, we define coherence $C$ by
$C =  2|\rho_{01}| = 2 |\rho_{10}|$,  a measure of  cross-terms in the
computational basis  not vanishing. The particle  is initially assumed
to be  in the state  $|\psi_a\rangle$, and a ``monitor",  initially in
the  state $|0\rangle$,  interacting with  each other  by means  of an
interaction $U$, parametrized by variable $\theta$ that determines the
entangling  strength  of  $U$.   It  is  convenient  to  choose  $U  =
\cos\theta~I\otimes I  + i\sin\theta {\rm  ~CNOT}$, where CNOT  is the
operation $I\otimes|0\rangle\langle0| + X \otimes |1\rangle\langle1|$,
where $X$  is the  Pauli $X$  operator. Under the  action of  $U$, the
system    particle    goes    to    the    state    $\rho    =    {\rm
  Tr}_m[U(|\psi_a\rangle|0\rangle\langle\psi_a|   \langle0|)U^\dag]  =
\frac{I}{2}   +  \frac{1}{2}[  (\cos\theta   +  i\sin\theta)\cos\theta
  |0\rangle\langle1|  +  {\rm  c.c}$, where  Tr$_m[\cdots]$  indicates
  taking  trace over  the  monitor.  Applying  the  above formula  for
  coherence to  $\rho$, we calculate that coherence  $C = \cos\theta$.
  We   let   $\lambda_{\pm}$  denote   the   eigenvalues  of   $\rho$.
  Quantifying the degree of entanglement by concurrence \cite{woo}, we
  have  $E \equiv 2\sqrt{\lambda_-\lambda_+}  = \sin\theta$.   We thus
  obtain a trade-off between  coherence and entanglement given by $C^2
  +  E^2  =  1$,   a  manifestation  of  the  complementarity  between
  single-particle and two-particle interference.

In  the context  of  the  proposed experiment,  this  could raise  the
following purported objection to our proposed signaling scheme: as the
experiment  happens  in  the  near-field  regime,  where  two-particle
correlations are strong, one would  expect that Bob should not find an
interference  pattern in his  singles counts.   Yet, contrary  to this
expectation,  Eq.   (\ref{ram})  implies  that  such  an  interference
pattern  does  appear.   The  reason   is  that  in  the  focal  plane
measurement,  Alice is  able  to  erase her  path  information in  the
subspace   ${\cal   H}_2$,   but,   by  virtue   of   the   associated
non-completeness, she  does so  in {\em only  one} way, viz.   via the
non-complete operation $E_f^{(+)}$ associated with her measurment.  If
her measurement were {\em  complete}, she would erase path information
in   more   than   one   way,  and   the   corresponding   conditional
single-particle interference patterns would mutually cancel each other
in the singles count. This is clarified in the following Section.
 
\subsection{Non-completeness implies lack of complementary measurement}

Let  us suppose  Alice's  measurement at  $f$  is replaced  by a  {\em
  complete} scheme  in which  her measurement is  deferred to  a point
behind a beam-splitter placed at  $f$, whereby the path singularity is
removed.  The action of such a beam splitter is
\begin{eqnarray}
\hat{a}_2      &\longrightarrow&      \hat{a}_{2^{\prime}}      \equiv
\cos\beta~\hat{a}_2     +    i\sin\beta~\hat{a}_5,     \\    \hat{a}_5
&\longrightarrow&  \hat{a}_{5^{\prime}} \equiv  i\sin\beta~\hat{a}_2 +
\cos\beta~\hat{a}_5.
\label{eq:sribhim}
\end{eqnarray}
Completeness               nows               holds              since
$\hat{a}^\dag_{2^\prime}\hat{a}_{2^\prime}                            +
\hat{a}^\dag_{5^\prime}\hat{a}_{5^\prime} =  \mathbb{I}_2$.  From Eqs.
(\ref{spdc}),    (\ref{Eq:coinc})     and    (\ref{bobfjeld}),    with
$\hat{a}_j^\prime$ replacing $\hat{a}_j$ ($j = 2,5$),
we find that  the joint probability for detection of  a photon in mode
$2^\prime$ or $5^\prime$ by Alice and at $z$ by Bob is given by
\begin{subequations}
\begin{eqnarray}
p_{2^\prime}(z)  &\propto&   \langle  E^\dag_z  a^\dag_{2^\prime}  E_z
a_{2^\prime}  \rangle = \frac{1}{2}(1  + \sin(2\beta)\sin[k(r_5-r_2)])
\\  p_{5^\prime}(z) &\propto&  \langle E^\dag_z  a^\dag_{5^\prime} E_z
a_{2^\prime} \rangle = \frac{1}{2}(1 - \sin(2\beta)\sin[k(r_5-r_2)]).
\end{eqnarray}
\end{subequations}
Tracing   over   Alice's   outcomes,   we  find   $p_{2^\prime}(z)   +
p_{5^\prime}(z) \propto 1$, and so no superluminal signaling occurs.

In  the light  of this,  let us  consider how  Alice's  momentum 
measurement could
seemingly  be  completed.  We  restrict  ourselves  to the  simplified
first-quantization  representation  of  the  field  state  vector  Eq.
(\ref{eq:srispdc}).   The measurement  operator  corresponding to  her
detection at  $f$ is  $\mathbb{P}_f \equiv E_f^{(-)}  E_f^{(+)} \equiv
\frac{1}{2}(|0\rangle + |1\rangle)(\langle0|  + \langle1|)$ in view of
Eq. (\ref{eq:srif0}).   One might consider that to  make her measurement
complete, $\mathbb{P}_f$ should be complemented by
\begin{equation}
\overline{\mathbb{P}}_f \equiv \mathbb{I}_2 - \mathbb{P}_f 
= 
\frac{1}{2}(|0\rangle - |1\rangle)(\langle0| - \langle1|),
\label{eq:srif01}
\end{equation}
which might  be interpreted as  the operator corresponding  to Alice's
non-detection at  $f$.  If Alice's momentum measurement  were given by
the  pair   $\{\mathbb{P}_f,  \overline{\mathbb{P}}_f\}$,  clearly  no
superluminal signal occurs for the reason given above.  Here one might
consider  $\overline{\mathbb{P}}_f$ to  the operator  corresponding to
Alice's   non-detection   at   $f$.    Unfortunately,   the   operator
$\overline{\mathbb{P}}_f$  in necessarily  non-physical, which  can be
seen in several ways.

Let  us  consider  what  $\overline{\mathbb{P}}_f$ represents  from  a
quantum  optics (second  quantization)  perspective.  Converting  from
first quantization  langauge, we see  that it represents  $\hat{a}_2 -
\hat{a}_5$. But this  does {\em not} correspond to  the electric field
operator at  {\em any} point on  Alice's side, since  these modes meet
only at  $f$, and since  $p$ and $q$  are equidistant from  $f$ (along
optical  rays), the  form of  the electric  field operator  at  $f$ is
$\hat{a}_2  + \hat{a}_5$,  which is  of course  $\mathbb{P}_f$  in the
first quantization langauge.   Thus $\overline{\mathbb{P}}_f$ is not a
valid measurement operator of Alice in the current set-up.

In  particular,  $\overline{\mathbb{P}}_f$   is  not  the  measurement
operator that corresponds  to her non-detection at $f$.  If Alice does
not detect her  photon at the $f$, then she  would in principle detect
them   elsewhere   on  the   focal   plane   (points  $f^\prime$   and
$f^{\prime\prime}$ in our present model of Figure \ref{fig:zei})).  In
our simplified 6-mode picture, they are given by the operators
\begin{eqnarray}
\mathbb{P}_{f^\prime} &=& \frac{1}{2}
(|1\rangle + e^{ik(r_{4f^\prime}-r_{1f^\prime})}|4\rangle)
(\langle 1| + e^{-ik(r_{4f^\prime}-r_{1f^\prime})}\langle4|) \nonumber \\
\mathbb{P}_{f^{\prime\prime}} &=& \frac{1}{2}
 (|3\rangle + e^{ik(r_{6f^{\prime\prime}}-r_{3f^{\prime\prime}})}|6\rangle)
(\langle 3| + e^{-ik(r_{6f^{\prime\prime}}-r_{3f^{\prime\prime}})}\langle6|),
\end{eqnarray}
in view of Eqs. (\ref{alicefjeldab}) and (\ref{alicefjeldac}).

But  clearly   $\overline{\mathbb{P}}_f  \ne  \mathbb{P}_{f^\prime}  +
\mathbb{P}_{f^{\prime\prime}}$.  In fact, the two operators don't even
have  the  same  support.   Thus  $\overline{\mathbb{P}}_f$  does  not
correspond to non-detection  at $f$ and could not  be used to complete
$\overline{\mathbb{P}}_f$.This   strange  state   of   affairs  is   a
consequence   of   non-completeness   as   had   been   clarified   in
note-in-citation \cite{sridir}. In other words, Alice's non-detection at
$f$ does not give rise to a complementary interference pattern, but
to a non-detection at Bob's side, too.

\subsection{Polarization and interference \label{sec:sriprim}}

The physical realization  of $Q$ allows us to  study the polynomiality
of the family of $Q$-like gates from a less abstract and more physical
perspective.  The gate  $Q_2(\pi)$ acting on $(1/\sqrt{2})(|0\rangle +
|1\rangle)$  annihilates it. Physically  this describes  the situation
where two  converging modes at  the path singularity, having  the {\em
same} polarization, interfere with each other destructively, resulting
in no particles being observed.  This is analogous to the situation of
dark fringes in a  Young's double-slit experiment.  The quantum optics
formalism implies that if the  polarizations of the two incoming modes
are not parallel, then the polarizations add vectorially (whether in a
complete or non-complete  configuration), with the resulting intensity
being the squared magnitude of the vector sum.

This point is worth stressing, since  if it were not so, it could give
rise  to  `interferometric quantum  computing'  that  would allow  for
efficient solution of hard problems.   For example, suppose we have an
$N  \equiv 2^n$  dimensional system  defined on  $n$  qubits, prepared
initially  in  the   state  $|a\rangle  \equiv  (1/\sqrt{N})(|1\rangle
+\cdots  + |N\rangle)$.  The  spatial part  of the  physical $n$-qubit
system's  matter  wave is  now  split into  two  partial  waves by  an
appropriate beam-splitter, and then refocused onto a path singularity.
On the  second partial  wave, before the  two partial waves  reach the
region of spatial  overlap, an oracle operation is  applied which in a
single  step inverts the  sign of  all the  kets, except  the `marked'
state $|N\rangle$,  yielding $|b\rangle \equiv (1/\sqrt{N})(-|1\rangle
-\cdots  +  |N\rangle)$.   According  to the  above  prescription  for
non-complete detection,  the output at the path  singularity should be
$|a\rangle  + |b\rangle  \sim 2|N\rangle/\sqrt{N}$,  i.e.,  an outcome
$|N\rangle$  observed   with  the  exponentially  low   (in  terms  of
$\log(N)$, the number of qubits used to realize the state) probability
of $|||a\rangle  + |b\rangle||^2 = 4/N$.   The physical interpretation
is that  for the most part,  with probability $(N-4)/N  \approx 1$ for
large  enough $n$,  no atoms  are  observed at  the path  singularity,
whereas the  solution state  $|N\rangle$ is observed  with probability
$4/N$.   Here   non-detection  of  atoms  should   be  interpreted  as
suppression of spatial transfer of atoms from their source to the path
singularity.   This is  reminiscent of  coherent  population trapping,
where an atom in a `dark state' remains unexcited because two pathways
to excitation destructively interfere \cite{scuzub}.

Therefore, if the above oracle  operation could be defined so that the
marked state  is a solution to  SAT, the measurement would  have to be
repeated  exponentially large  number of  times to  detect  a possible
`yes'  outcome.  Alternatively, exponentially  large  number of  atoms
should be  used to  build up  an answer signal  of strength  $O(1)$ in
polynomial  time.  Either  way, the  physical situation  is compatible
with the WNHE assumption, but not with no-signaling.

The implementation  of non-complete measurement of  modes of arbitrary
polarization  gives  us  further  insight into  the  polynomiality  of
Nature.  It is  not difficult to imagine a  more complicated rule than
plain vector  addition for  the interference of  quantum wavefunctions
(say a  renormalization following  vector addition), which  could have
been used to boost the above  signal, to solve SAT in polynomial time.
This  would in fact  implement the  post-selection gate.   However, it
would have  required `Nature  to compute harder'  than believed  to be
possible with a Turing machine  or equivalent model of computation, in
contradiction to the WNHE viewpoint  that the universe is a polynomial
place.

\section{Discussions and Conclusions}\label{sec:conclu} 

Considering the far-reaching implications  of a positive result to our
proposed experiment, we  have to remain open to  the possibility that
there  is  an error  somewhere  in  our  analysis, possibly  a  hidden
unwarranted assumption, the elimination of which would provide a
mechanism to prevent the superluminal signaling.  
However,  in support of our claim,  it may be
noted that our analysis of  the Modified Innsbruck experiment is based
on a  model that  works quite  well in explaining  the results  of the
original  Innsbruck experiment. This suggests that
it would be difficult to prohibit the superluminal signal in
the model of the Modified experiment without also ending up proscribing
two-particle correlations in the model for the Innsbruck experiment.

  Furthermore,  we have  ruled out  in
Section \ref{sec:sriung} all the  (so far as known to us)
obvious objections. Hence we are led
to believe  that any erroneous  assumption or application  of physical
principles, if it exists in our analysis, must be sufficiently subtle.
Therefore, it  would still be  instructive to perform  our proposed
experimental tests because, even if the tests yield a negative result,
provided the outcome is  unambiguous, we could re-examine our analysis
confident  of detecting  an erroneous  element that  is  otherwise not
obvious.  As  in  the  earlier  mentioned example  of  the  no-cloning
theorem,  even  this  potentially  negative  result  could  carry  new
theoretical insights.

On the other hand, in the surprising event the proposed
experiment  yields  a positive  outcome,  a  number  of issues  would
clearly be brought up.  Foremost among them: the apparent violation of
locality in standard, linear QM  would now emerge as a real violation,
and no-signaling would  no longer be a universal  condition. The issue
of  `speed  of   quantum  information'  \cite{srisal08}  would  assume
practical significance.


A putative positive outcome to either of the proposed experiment would
also  bolster the case  for believing  that the  WNHE assumption  is a
basic  principle of quantum  physics, while  undermining the  case for
no-signaling in QM.  It would  then follow that intractability, and by
extension  uncomputability, matter  to physics  in a  fundamental way.
This   would   suggest   that   physical  reality   is   fundamentally
computational in nature  \cite{sri06}. With this abstraction, physical
space  would be  regarded  as  a type  of  information, with  physical
separation no  genuine obstacle to  rapid communication in the  way it
seems  to  be    when  seen  from  the  conventional  perspective   of  
causality  in physics.   On  the   other  hand,  the  barrier  between
polynomial-time  and  hard  problems  would  be  real.   The  physical
existence of superluminal  signals would thus not be  as surprising as
that  of exponential  gates.   Interestingly, polynomial  superluminal
operations  exist even  in classical  computation.  The  Random Access
Machine (RAM) model \cite{ste09}, a standard model in computer science
wherein memory access takes  exactly one time-step irrespective of the
physical location of the  memory element, illustrates this idea.  RAMs
are known  to be polynomially  equivalent to Turing machines.   At the
least,  WNHE  could  serve as  an  informal  guide  to issues  in  the
foundations of QM, and perhaps even quantum gravity.


Even granting  that the  noncomplete gate $Q^\prime$  turns out  to be
physically valid  and realizable, this brings us  to another important
issue: how  would non-completeness fit in with  the known mathematical
structure of the quantum properties  of particles and fields, and why,
if  true,  should it  have  remained  theoretically  unnoticed so  far
inspite of its far-reaching  consequences?  We venture that the answer
has to do  with the nature of and  relationship between observables in
QM on the  one hand, and those in quantum  optics, and more generally,
in quantum field theory (QFT), on the other hand.

It is frequently claimed that QFT  is just the standard rules of first
quantization  applied to classical  fields, but  this position  can be
criticized    \cite{srizeh,srihrv0,srihrv}.     For    example,    the
relativistic  effects   of  the   integer-spin  QFT  imply   that  the
wavefunctions describing a fixed number  of particles do not admit the
usual  probabilistic interpretation \cite{srihrv0}.   Again, fermionic
fields do not really have a classical counterpart and do not represent
quantum observables \cite{srihrv}.

In practice, measurable properties resulting from a QFT are properties
of particles--  of photons in quantum  optics.  Particulate properties
such  as number,  described by  the number  operator  constructed from
fields,  or the momentum  operator, which  allows the  reproduction of
single-particle QM in  momentum space, do not present  a problem.  The
problem is  the {\em position} variable,  which is considered  to be a
parameter,   and  not   a  Hermitian   operator,  both   in   QFT  and
single-particle relativistic QM,  and yet relevant experiments measure
particle positions.   The experiment described in  this work involve
measurement  of the  positions  of photons,  as  for example,  Alice's
detection of photons at points on the imaging or focal plane, or Bob's
detection at points on the $z$-plane, respectively.  There seems to be
no way to derive from  QFT the experimentally confirmed Born rule that
the nonrelativistic wavefunction  $\psi({\bf x},t)$ determines quantum
probabilities  $|\psi({\bf x},t)|^2$ of  particle positions.   In most
practical situations, this is really not a problem.  The probabilities
in the  above experiment were computed according  to standard quantum
optical rules to determine the correlation functions at various orders
\cite{glauber},  which  serve  as  an effective  wavefunction  of  the
photon,  as seen  for example  from Eqs.   (\ref{Eq:coinc}).   In QFT,
particle  physics phenomenologists have  developed intuitive  rules to
predict distributions of particle positions from scattering amplitudes
in {\em momentum} space.

Nevertheless, there  is a  problem in principle,  and leads us  to ask
whether QFT is  a genuine quantum theory \cite{srihrv}.   If we accept
that properties like position are  valid observables in QM, the answer
seems to  be `no'.  We see this  again in the fact  that the effective
'momentum'  and  'position'  observables   that  arise  in  the  above
experiment are not seen to be Hermitian operators of standard QM (cf.
note   \cite{sridir}).     Further,   non-complete   operations   like
$\hat{E}^{(+)}_f$,  disallowed in  QM, seem  to appear  in  QFT.  This
suggests that  it is QM,  and not QFT,  that is proved to  be strictly
non-signaling by the no-signaling theorem.

Since nonrelativistic  QM and QFT  are presumably not  two independent
theories describing  entirely different  objects, but do  describe the
same   particles  in   many  situations,   the   relationship  between
observables  in  the  two  theories  needs to  be  better  understood.
Perhaps some  quantum mechanical observables are  a coarse-graining of
QFT  ones,  having wide  but  not  universal  validity.  For  example,
Alice's  detection of  a photon  at  a point  in the  focal plane  was
quantum mechanically  understood to project the state  of Bob's photon
into   a  one-dimensional   subspace  corresponding   to   a  momentum
eigenstate.    Quantum  optically,   however,  this   `eigenstate'  is
described as a  superposition of a number of  parallel, in-phase modes
originating  from different down-conversion  events in  the non-linear
crystal, producing  a coherent plane wave propagating  in a particular
direction.

\acknowledgments  I  thank  the  anonymous  Referee,  whose  important
comments  have helped improve  the manuscript.  I am thankful to  Prof.  S.
Rangwala and  Mr. K. Ravi  for insightful elucidation  of experimental
issues,  and  to  Prof.  J.   Sarfatti, Prof.   J.  Cramer  and  Prof.
A. Shiekh for helpful discussions.


\begin{thebibliography}{100}
\bibitem{epr} A. Einstein, N. Rosen, and B. Podolsky.
Is the Quantum-Mechanical Description of Physical Reality Complete?
Phys. Rev. Lett. {\bf 47}, 777 (1935).
\bibitem{nos}  P.   H.  Eberhard,  Nuovo  Cimento   46B,  392  (1978);
C. D.  Cantrell and M.  O. Scully, Phys.  Rep.  {\bf 43},  499 (1978).
G. C. Ghirardi, A. Rimini, and T. Weber, Lett. Nuovo Cimento {\bf 27},
293  (1980).   P.  J.  Bussey,  Phys.  Lett.  {\bf  A  90},  9  (1982).
T. F.  Jordan, Phys. Lett. {\bf  94}A, 264 (1983).  A.  J. M. Garrett,
Found. Phys. {\bf  20} 381 (1990);  
A. Shimony, in {\em Proc. of
the Int. Symp. on Foundations  of Quantum Mech.}, eds. S. Kamefuchi et
al. (Phys.  Soc. Japan, Tokyo, 1993);   R Srikanth, Phys  Lett. {\bf A
292}, 161 (2002).
\bibitem{bell} J. S. Bell, Physics {\bf 1}, 195 (1964); J. F. Clauser,
M. A.  Horne, A. Shimony, and R.  A. Holt, Phys. Rev.  Lett. {\bf 23},
880 (1969); A.  Aspect, P.  Grangier and G.  Roger, Phys.  Rev.  Lett.
{\bf  49}, 91  (1982); W.   Tittel, J.   Brendel, H.   Zbinden  and N.
Gisin,  Phys.  Rev.   Lett  {\bf   81}  3563  (1998);  G.   Weihs,  T.
Jennewein,   C.     Simon,   H.   Weinfurter    and   A.    Zeilinger,
Phys. Rev. Lett.  {\bf 81}, 5039 (1998); P. Werbos, arXiv:0801.1234.
\bibitem{gle57} A. M.  Gleason. Measures on the closed  subspaces of a
Hilbert space.  J. Math. Mech., {\bf 6}, 885 (1957).
\bibitem{ash95} A. Peres, {\em Quantum Mechanics: Concepts and Methods},
(Kluwer, Dordrecht, 1993).
\bibitem{gis01halv}  C.    Simon,  V.   Bu{\v  z}ek   and  N.   Gisin,
Phys. Rev. lett. {\bf 87}, 170405 (2001); {\em ibid.} {\bf 90}, 208902
(2003).  H.   Halvorson, Studies in  History and Philosophy  of Modern
Physics 35, 277 (2004); quant-ph/0310101.
\bibitem{bra06} G. Brassard, H. Buhrman  and N. Linden et al.  A limit
on nonlocality in  any world in which communication  complexity is not
trivial.  Phys. Rev. Lett. {\bf 96} 250401 (2006).
\bibitem{sri06} R. Srikanth.
The quantum measurement problem and physical reality: a computation 
theoretic perspective.  AIP   Conference  Proceedings {\bf 864},
(Ed. D. Goswami) 178 (2006); quant-ph/0602114v2.
\bibitem{srirp}  In  complexity  theory,  {\bf  RP} is  the  class  of
decision  problems  for  which  there  exists a  probabilistic  TM  (a
deterministic TM with access to genuine randomness) such that: it runs
in  polynomial time  in the  input  size. If  the answer  is `no',  it
returns  `no'. If  the  answer   is  `yes',  it  returns  `yes'  with
probability at least 0.5 (else it returns `no').
{\bf  BQP} is
the  class  of  decision  problems  solvable by  a  {\em  quantum}  TM
\cite{nc00} in polynomial time, with  error probability of at most 1/3
(or,  equivalently,  any  other   fixed  fraction  smaller  than  1/2)
independently of input size.
\bibitem{nc00} M. A. Nielsen and I. Chuang, 
{\it Quantum Computation and Quantum Information}, (Cambridge 2000).
\bibitem{wit89} E. Witten. Quantum field theory and the Jones polynomial. 
Commun. Math. Phys. {\bf 121}, 351 (1989).
\bibitem{mic98} M.   Freedman. {\em P}/{\em PN} and  the quantum field
computer.  Proc. Natl. Acad. Sci. USA {\bf 95}, 98 (1998).
\bibitem{cal04} C. S.  Calude, M.   A. Stay. From Heisenberg to Goedel
via Chaitin.  Int.  J.  Th.  Phys.  44 1053 (2005); quant-ph/0402197.
\bibitem{insel} S.  Aaronson,
quant-ph/0401062; {\em ibid.},
quant-ph/0412187.
\bibitem{aar05} S. Aaronson. NP-complete Problems and Physical Reality.
ACM SIGACT News {\bf 36} 30 (2005); quant-ph/0502072. 
\bibitem{bram98}  D.  S  Abrams   and  S. Lloyd.   Nonlinear  quantum
mechanics  implies polynomial-time  solution for  NP-complete  and \#P
problems. Phys. Rev. Lett. {\bf 81}, 3992 (1998).
\bibitem{sciam} S. Aaronson. The limits of quantum computers.
American Scientist {\bf 42}, March 2008.
\bibitem{srigru}  J. Gruska. Quantum  informatics paradigms  and tools
for QIPC.  BackAction Quantum Computing: Back Action 2006, IIT Kanpur,
India,   March   2006,   Ed.    Dr.   D.   Goswami,   AIP   Conference
Proceedings. 864, pp. 178193 (2006).
\bibitem{sriwnhe} That  is, `the universe  is not hard enough  to {\em
  not} be  simulable using  polynomial resources'.  The  expression is
  non-technically related to the  statement ``The world is not enough"
  (``{\em orbis  non sufficit}''), the family  motto of, as  well as a
  motion picture featuring, a well known Anglo-Scottish secret agent!
\bibitem{sriw}  S. Weinberg.  {\em  Dreams of  a Final  Theory}
(Vintage 1994).
\bibitem{sribpp}  More formally, {\bf  BPP} is  the class  of decision
problems solvable by a probabilistic TM in polynomial time, with error
probability of at most 1/3 (or, equivalently, any other fixed fraction
smaller than 1/2) independently of input size.
\bibitem{bag00} A. Bassi and  G. Ghirardi.  A General Argument Against
the Universal Validity of  the Superposition Principle.  Phys. Lett. A
275 (2000); quant-ph/0009020.
\bibitem{sriv1} R. Srikanth. No-signaling, intractability and
entanglement. Eprint 0809.0600v1.
\bibitem{ganeshpol91}  N.  Gisin, Helv.  Physica  Acta  {\bf  62}, 383  (1989);
Phys. Lett. {\bf A} 143, 1  (1990). 
J.  Polchinski.  
Phys. Rev. Lett. {\bf 66}, 397 (1991).
\bibitem{srisvet} G. Svetlichny, Int. J. Theor. Phys. {\bf 44},
2051 (2005): quant-ph/0410230. 
\bibitem{srisvet0} G. Svetlichny. Informal Resource Letter -- 
Nonlinear quantum mechanics. quant-ph/0410036.
\bibitem{sen} A.  Sen-De and U.  Sen.  Testing quantum  dynamics using
signaling. Phys. Rev. A {\bf 72}, 014304 (2005).
\bibitem{shiekh}  A.  Shiekh.  Quantum Interference:  an  experimental
proposal, and possible Engineering applications. arXiv:0901.1475.
\bibitem{pspace}  {\bf  PSPACE}  is  the class  of  decision  problems
  solvable by  a Turing  machine in polynomial  (memory) space possibly 
taking exponential time.
\bibitem{sripp}  In  complexity  theory,  {\bf  PP} is  the  class  of
decision   problems  for   which  there   exists  a   polynomial  time
probabilistic TM such  that: if the answer is  `yes', it returns `yes'
with probability greater than $1/2$, and if the answer is `no', it
returns `yes' with probability at most $1/2$.  
\bibitem{srigro97} Grover, L. K. Quantum mechanics helps in searching for
a needle in a haystack.  Phys. Rev. Lett. {\bf 79}, 325 (1997).
\bibitem{sriben97}   C.  Bennett et al.
Strengths     and     weaknesses    of     quantum
computing. quant-ph/9701001.
\bibitem{sriper02} A. Peres. How the no-cloning theorem got its name.
quant-ph/0205076.
\bibitem{srinom}  
Operations
(\ref{eq:Kjanichwara})  are in  fact  a kind  of superquantum  deleter
envisaged in Ref. \cite{sen}.
\bibitem{sri07} R.  Srikanth and S.  Banerjee. An environment-mediated
quantum deleter. Phys. Lett. A {\bf 367}, 295 (2007); quant-ph/0611161.
\bibitem{asp8} R. Ghosh and L. Mandel, Phys. Rev. Lett. {\bf 59}, 1903
(1987); P. G. Kwiat,  A. M. Steinberg and R. Y. Chiao,  Phys. Rev. {\bf A
47}, 2472 (1993); D. V. Strekalov, A. V. Sergienko, D. N. Klyshko, and
Y. H.  Shih, Phys. Rev.  Lett. {\bf 74},  3600 (1995); T.  B. Pittman,
Y. H.  Shih, D. V. Strekalov, and  A. V. Sergienko, Phys.  Rev. {\bf A
52},  R3429 (1995);  Y. -H.  Kim, R.  Yu, S.  P. Kulik,  and  Y. Shih,
M. O. Scully, Phys. Rev. Lett. {\bf 84}, 1 (2000).
\bibitem{zei00} A. Zeilinger. Experiment and foundations of quantum
physics. Rev. Mod. Phys. {\bf 71}, S288 (1999).
\bibitem{dop98}  B.  Dopfer.  Zwei  Experimente  zur  Interferenz  von
  Zwei-Photonen     Zust\"anden:    ein     Heisenbergmikroskop    und
  Pendell\"osung.  Ph.D. thesis (University of Innsbruck, 1998).
\bibitem{sri} R. Srikanth, Pramana {\bf 59}, 169 (2002); 
 {\em ibid.} quant-ph/0101023. 
\bibitem{sri888}   R.    Srikanth.    quant-ph/9904075;  {\em   ibid.}
  quant-ph/0101022.
\bibitem{sricram} J. G. Cramer, W. G. Nagourney and S. Mzali.
A test of quantum nonlocal communication. CENPA Annual Report (2007); \\
http://www.npl.washington.edu/npl/int\_rep/qm\_nl.html and
http://faculty.washington.edu/jcramer/NLS/NL\_signal.htm.
\bibitem{sridom}  R.   Jensen,  STAIF-2006  Proc.,   M.  El-Genk,
ed.                  1409                 (AIP,                 2006);
http://casimirinstitute.net/coherence/Jensen.pdf (2006).
\bibitem{srihrv} H. Nikoli\v{c}.
Is quantum field theory a genuine quantum theory? Foundational insights 
on particles and strings.  arXiv:0705.3542.
\bibitem{peres} A. Peres, FLASH (2003).
\bibitem{ganeshchandru} S. Banerjee, C. M. Chandrasekhar, R. Srikanth,
in preparation.
\bibitem{sridir}  We  can   then  define  Alice's  (remote)  `momentum
  observable'  as   \mbox{$\hat{P}  \equiv  f|2+5\rangle\langle2+5|  +
    f^{\prime}|1+4\rangle\langle1+4|                                  +
    f^{\prime\prime}|3+6\rangle\langle3+6|$}.     Interpreted   as   a
  quantum  field  theoretic   observable,  $\hat{P}$  is  non-complete
  because   the   projectors   to   its   eigenstates   $|2+5\rangle$,
  $|1+4\rangle$ and  $|3+6\rangle$ do not sum  to $\mathbb{I}_6 \equiv
  |1\rangle\langle1|  +   |2\rangle\langle2|  +  |3\rangle\langle3|  +
  |4\rangle\langle4|   +  |5\rangle\langle5|   +  |6\rangle\langle6|$.
  Similarly,  Alice's  non-complete  `position' observable  is  \mbox{
    $\hat{Q}        \equiv        m|1+2+3\rangle\langle1+2+3|        +
    l|4+5+6\rangle\langle4+5+6|$}.    But    note   that   $\hat{Q}$'s
  projection into  the subspace ${\cal H}_{2,5}$ is  indeed a complete
  observable.   $\hat{P}$   and  $\hat{Q}$  are  of  rank   3  and  2,
  respectively, which is smaller than 6, the dimension of the relevant
  Hilbert subspace.
\bibitem{abu01} A.F. Abouraddy, et al.
Demonstration of the complementarity of one- and two-photon interference.
Phys. Rev. A {\bf 63}, 063803 (2001).
\bibitem{bos02} S. Bose and D. Home.
Generic Entanglement Generation, Quantum Statistics, and Complementarity.
 Phys. Rev. Lett. {\bf 88}, 050401 (2002).
\bibitem{scuzub} M. Scully and S. Zubairy.
{\em Quantum Optics} (Cambridge University Press, 1997).
\bibitem{srisal08} D. Salart, A. Baas, C. Branciard, N. Gisin, and H. Zbinden.
Testing the speed of `spooky action at a distance'.
Nature 454, 861 (2008); arxiv:0808.331v1.
\bibitem{ste09} S. Skiena, {\em The Algorithm Design Manuel},
Springer (1998).
\bibitem{srizeh} H. D. Zeh. There is no "first" quantization.
Phys. Lett. A {\bf 309}, 329 (2003); quant-ph/0210098.
\bibitem{srihrv0} H. Nikoli\v{c}. There is no first quantization - 
except in the de Broglie-Bohm interpretation. quant-ph/0307179.
\bibitem{glauber} R. J. Glauber.
The Quantum Theory of Optical Coherence. Phys. Rev. 130, 2529 (1963).
\bibitem{woo} W. K. Wootters. Entanglement of Formation and Concurrence.
Quant. Info. and Comput. {\bf 1}, 27 (2001).
\end{thebibliography}
\end{document}